\documentclass[a4paper,fleqn,usenatbib]{mnras}

\usepackage[T1]{fontenc}
\usepackage{ae,aecompl}

\usepackage{graphicx}	
\usepackage{amsmath}	
\usepackage{amssymb}	

\usepackage{rotating}
\usepackage{threeparttable}

\usepackage{dcolumn}
\usepackage{longtable}
\usepackage{multirow}
\usepackage{bm}
\usepackage{textpos}
\usepackage{pdflscape}	

\usepackage{geometry}
\usepackage{tikz}
\usepackage{ifthen}
\usetikzlibrary{calc}

\title[Stellar Triples as a Source for Ba Stars]{Stellar Triples as a Source for Ba Stars}
\author[Y. Gao et. al.]{
Yan Gao,$^{1}$\thanks{E-mail: ygbcyy@star.sr.bham.ac.uk}
Silvia Toonen,$^{2}$
Nathan Leigh$^{3,4}$
\\
$^{1}$Institute for Gravitational Wave Astronomy \& School of Physics and Astronomy, \\
University of Birmingham, Edgbaston, Birmingham B15 2TT, UK\\
$^{2}$Anton Pannekoek Institute for Astronomy, University of Amsterdam, 1090 GE Amsterdam, The Netherlands\\
$^{3}$Departamento de Astronom\'ia, Facultad Ciencias F\'isicas y Matem\'aticas,
Universidad de Concepci\'on, \\
Av. Esteban Iturra s/n Barrio Universitario,\\ Casilla
160-C, Concepci\'on, Chile \\
$^{4}$Department of Astrophysics, American Museum of Natural History, New York, NY
10024, USA \\
}

\date{Accepted XXX. Received YYY; in original form ZZZ}

\pubyear{2022}

\begin{document}
\label{firstpage}
\pagerange{\pageref{firstpage}--\pageref{lastpage}}
\maketitle

\begin{abstract}
Barium stars have been studied extensively over the past few decades, yet our current understanding of how these intriguing objects formed leaves much to be desired. Many trends observed in systems containing barium stars cannot be satisfactorily explained by classical binary evolution models, naturally raising the question of whether triples and other higher-order multiples can give rise to such exotic objects. In this paper, we study the possibility that a Roche Lobe overflow from a tertiary in a hierarchical triple system can potentially lead to surface barium enrichment within the inner binary, while at the same time causing the inner binary to merge, thereby producing a barium star. This possibility has the potential to form a large proportion of Barium stars, as Roche Lobe overflow from a tertiary is typically much more stable for close orbits than that from a binary companion. Various formation channels and mechanisms by which this can be achieved are considered, and constraints on relative formation rates are placed on each scenario. We conclude that a significant, if not dominant, proportion of barium stars are formed from hierarchical triple systems, and that further studies are required in this area before a complete understanding of Barium star populations can be achieved.
\end{abstract}





\section{Introduction}


Usually, barium (Ba) in stars is a slow neutron capture process (s-process) product that is only found in the later stages of stellar evolution. This is because the prerequisite s-process that synthesises this element only starts to occur during the asymptotic giant branch (AGB) phase of stellar evolution, which only takes place long after the main sequence (MS) and red giant branch (RGB) phases have ended. However, many pre-AGB stars have been demonstrated to have detectable traces of Ba at their stellar surfaces \citep{1951ApJ...114..473B}, and astronomers were quick to define these irregular objects as ``barium stars" \citep[e.g.][]{1965MNRAS.129..263W}.

Over the years, much has been uncovered regarding the origins of the chemical peculiarity of these Ba stars. Their counterintuitively high surface Ba content turned out to be due to donations from a stellar companion \citep[e.g.][]{1988A&A...205..155B}. Once the companion evolves into an AGB star, it will be producing Ba, and will at the same time be expanding in size. If this companion is in a close enough orbit, then it will eventually fill its Roche Lobe, and transfer its mass to the primary, endowing the primary with matter that has the same chemical composition as its own outer layers, and enriching it with Ba through this process. Since the simplest conceivable stellar system that allows such a process to occur is a binary system, application of Occam's razor leads naturally to a standing consensus that Ba star progenitors are binary systems. 


However, recent work has discovered inconsistencies between the statistical orbital properties of binaries containing a Ba star and what is predicted by theory, assuming that all Ba star progenitors are binary systems. For instance, when Ba-star-producing binaries evolve into their characteristic AGB phase, the subsequent mass transfer from the more massive star to its less massive companion will induce unstable mass transfer, which will in turn lead to either tighter orbits with little accretion or a merger event. This leads to a range of orbital periods around the 1000 day mark in which Ba stars should not be found, whereas in reality there are many Ba stars co-orbiting companions in this period range \citep{2010A&A...523A..10I}. Also problematic is the issue that current theories predict that binaries containing Ba stars with orbital periods less than 4000 days should be highly circular, whereas in reality such systems frequently have large eccentricities, sometimes up to 0.4 \citep[e.g.][]{1998A&A...332..877J,2019A&A...626A.128E,2020Obs...140...11N}. 

To address these problems, many solutions have been proposed to reconcile Ba star progenitor models with observations in the context of binary evolution. In answer to the issue that long-period Ba star binaries can have high eccentricities, it has been proposed that some peculiarity of the mass transfer from the AGB to the Ba star may lead to the eccentricity being retained. Many such eccentricity-retention mechanisms have been studied, such as the periodic mass loss rates in resonance with the orbit \citep[e.g.][]{2008A&A...480..797B}, and interaction with a circumbinary disk from which the mass is being accreted (e.g. \citealt{2013A&A...551A..50D,2015A&A...579A..49V}, see also \citealt{2016ApJ...830....8R}). However, despite these solutions being able to account for the survival of large eccentricities during and hence immediately after mass transfer, none of them are capable of explaining how the eccentricity can be maintained long after the mass transfer has ended, and there is evidence that Ba giants, which have evolved for a long time after receiving the initial Ba injection from their companions, can have highly eccentric orbits \citep{2020A&A...639A..24E}. As for the existence of Ba stars with orbital periods shorter than 1000 days, \cite{2020A&A...639A..24E} investigated the possibility that Ba stars can themselves undergo a giant phase and interact with their companion, resulting in a different orbital separation, and found that this could produce Ba star binaries in the 1000-day range. However, they note that this does not reproduce the short-period end of the observed Ba giant population, and also point out that this would result in eccentricities far in excess of what is seen. In short, these solutions cannot fully account for observations. In the absence of formation channels that have not yet been thought of, explaining these discrepancies would be a daunting task indeed.

One possible way to overcome this quandary is to propose that a significant proportion of Ba stars do not form merely as a result of binary interactions alone. If this proposition is true, then one probable candidate for Ba star creation would be triple stellar systems. We already know that triple systems constitute 13\% of all stellar systems consisting of low-mass stars \citep[e.g.][]{2014AJ....147...87T}, or in other words, 1 in 3 binary systems have a tertiary companion. This ratio is drastically higher for stars of higher masses \citep{2012Sci...337..444S,2017IAUS..329..110S,2017ApJS..230...15M}, the average multiplicity of which has been found to be 3. 

Given their prevalence, it should not be hard to imagine that interacting triple systems can lead to Ba stars. Given their structural complexity, it should not be hard to imagine that they can lead to Ba stars in more ways than one. The existence of a tertiary can potentially drive a previously non-interacting inner binary to exchange mass during the AGB phase of one of its components \citep{2020A&A...640A..16T}. Alternatively, an AGB-phase common envelope (CE) occurring within the inner binary can potentially eject Ba-rich material, which can then be accreted by the tertiary, and so the list continues. Among all these possibilities, which all doubtlessly deserve to be examined in future studies, we note that one possibility is particularly interesting, that in which the tertiary expands as an AGB star prior to the inner binary, and transfers Ba-rich material into the inner binary. Should the tertiary's influence be sufficient to merge the inner binary henceforth, the system will ultimately evolve into a binary in which the merged binary is a Ba star, and the tertiary will take on the form of a binary companion.

In this scenario, the combined masses of the inner binary can be greater than that of the tertiary, despite being individually less massive. If this is the case, the first mass transfer phase of the system, initiated by the tertiary, causes the orbital separation of the outer orbit to increase, thus avoiding the runaway mass transfer that is typical of analogous phases in binary evolution. We posit that this leads to this formation channel being doubly interesting, as it can give rise to behaviours within the system for which there is no counterpart in binary evolution. Furthermore, due to this increased level of stability following mass transfer, which binaries lack, it would not be surprising if a far greater proportion of  hierarchical triple systems which undergo mass transfer in this way lead to Ba stars than their binary counterparts. Thus, the rates at which Ba stars are formed from this channel could be on par, if not greater, than that of traditional binary formation channels.

In this paper, we seek to better understand the nuances of this potential Ba star formation channel in the context of the general triple population. To do this, we examine a population of triples generated via triple population synthesis, and constrain the number of Ba stars generated by means of a series of simple assumptions, followed by the corresponding calculations. This paper is divided into 5 sections, of which this introduction is the first. In the second section, we will summarise the possible ways by which tertiary Roche Lobe overflow (RLOF) can result in a Ba star, and provide the relevant prescriptions we adopt for each possible case. In the third section, we present our simulated data set and how we use it to place constraints on the prevalence rate of triple-origin Ba stars, the results of which are provided in the fourth section. Finally, in the fifth and final section, we will discuss the implications of what we arrived at in the broader context of stellar systems in general.

\section{Formation Channels and Mechanisms}

Consider a hierarchical triple, in which the inner binary consists of two bodies with masses $m_{\rm 1}$ and $m_{\rm 2}$, while the tertiary has a mass of $m_{\rm 3}$. For simplicity, we assume that $m_{\rm 2}<m_{\rm 1}$. The semi-major axes of the inner and outer orbits are $a_{\rm 1}$ and $a_{\rm 2}$ respectively, while the eccentricities are $e_{\rm 1}$ and $e_{\rm 2}$. The mass ratios of the inner and outer orbits, denoted as $q_{\rm 1}$ and $q_{\rm 2}$ respectively, are defined as $q_{\rm 1}=m_{\rm 2}/m_{\rm 1}$ and $q_{\rm 2}=m_{\rm 3}/(m_{\rm 1}+m_{\rm 2})$. The initial masses satisfy $m_{\rm 3}>m_{\rm 1}$ and $m_{\rm 3}>m_{\rm 2}$, and both the inner and outer orbital separations are small enough to allow Roche Lobe overflow. As such, the tertiary will evolve past the main sequence phase and fill its Roche Lobe before either of the inner binary components, resulting in the configuration we seek to examine.

\subsection{Mass Transfer from a Tertiary to an Inner Binary}

Upon filling its Roche Lobe, the tertiary will begin transferring mass to the inner binary. Depending on the specific circumstances, this can lead to either of two results: either unstable mass transfer ensues, and the transferred material eventually forms a CE around both the inner binary and the tertiary, or mass transfer is stable, and it does not form this circumtriple CE. In the latter case, a CE may or may not form around the inner binary, but we do not make this distinction for now. The watershed between these two possibilities (circumtriple CE or no circumtriple CE) is when 
\begin{equation}
2.13q_{\rm 2}=1.37+2\left(\frac{m_{\rm c}}{m_{\rm 3}}\right)^{\rm 5};
\label{outCEcrit}
\end{equation}
\noindent where $q_{\rm 2}{\equiv}m_{\rm 3}/\left(m_{\rm 1}+m_{\rm 2}\right)$, and $m_{\rm c}$ is the value of $m_{\rm 3}$ after the RLOF is over \citep{2002MNRAS.329..897H}. If $q_{\rm 2}$ is greater than that in Eq. \ref{outCEcrit}, a CE forms; if it is smaller, no CE forms.

For the case in which this circumtriple common envelope is formed, there is currently no easy presciption by which we can calculate what happens next; the question of what happens when circumtriple CEs are formed is still a heatedly contested debate, with predictions ranging from the merging of the inner binary \citep[e.g.][]{2021MNRAS.500.1921G} to outright disruption of the triple in question \citep{2020MNRAS.498.2957C}. Therefore, we concentrate our efforts on the case in which such a circumtriple CE does not form, and stable mass transfer ensues from the tertiary onto the inner binary.

\subsection{Comments on Potential Formation of a Disk}

From here, the matter flowing through the L1 Lagrange point will form a stream, which will flow towards the inner binary, and eventually enter into a Keplerian trajectory with the centre of mass (COM) of the inner binary. 

If the tip of the stream does not interact with the inner binary in any meaningful way, it will proceed to pass the point of periapsis and make a full orbit around the inner binary, whereupon it will strike another part of the same stream at the point where it first began its Keplerian orbit. This will have a tendency to push the trajectory of further infalling material outwards, away from the binary in relation to the original Keplerian orbit, ultimately resulting in a circular circumbinary disk with a radius \citep{2002apa..book.....F} of
\begin{equation}
r_{\rm disk}=a_{\rm 2}\left(1+q_{\rm 2}\right)\left(0.5-0.227{\log}q_{\rm 2}\right)^{\rm 4},
\label{diskcircrad1}
\end{equation}
or, according to \cite{2020MNRAS.496.1819L},
\begin{equation}
r_{\rm disk}=a_{\rm 2}\left(1-R_{\rm L}\right)^{\rm 2},
\label{diskcircrad2}
\end{equation}
\noindent where $R_{\rm L}$ is the ratio between the Roche Lobe radius of the tertiary and the outer orbital separation, as given by \cite{1983ApJ...268..368E}. 


On the other hand, if the inner binary is situated in a way that it obstructs the tip of the stream, a violent interaction will occur close to or when the tip is at the periapsis of the Keplerian orbit, and the tip of the stream is thus unable to make a full orbit and strike the rest of the stream, thus no disk can form. Here, it is interesting to note that the distance between the tip of the stream and the inner binary COM at periapsis is much smaller than the circularisation radius given by Eq. \ref{diskcircrad1} or \ref{diskcircrad2}. The upshot of this is that having an inner binary that is more compact than the circumbinary disk radius does not necessarily guarantee a disk; to argue that a disk will form, one must demonstrate that the inner binary orbital separation must not extend either binary component to the position of the periapsis of the initial orbit of the stream's tip. There are many ways by which various authors have attempted to make this distinction in the past, including but not limited to \cite{2020A&A...640A..16T}, who make the simplifying assumption that the periapsis is a constant fraction of $r_{\rm disk}$ away from the inner binary's COM. Regardless of how this is done, the evolutionary result must fall in either of two cases: one in which no disk forms, and another in which it does. We discuss the two cases in the following two subsections respectively.

\subsection{No Disk Forms}

In the case in which no disk forms, the subsequent evolution will ensure that a certain proportion between 0 and 1 of the mass in the disk must be accreted onto the inner binary, and the inner binary would then attempt to eject the remaining mass. Here we only need to consider two extreme regimes in order to place limits on what happens next; all other possibilities must lie in between. These two extreme regimes are the regime in which all the material is accreted, and the other in which all the material is ejected.

\subsubsection{Complete Accretion - No Disk}

In the event that no disk is formed, the infalling stream of material from the tertiary is disrupted by the inner binary. Extensive hydrodynamical simulations of the inner binaries of HD97131 and ${\xi}$ Tau when presented with a similar situation by \citealt{2014MNRAS.438.1909D} have shown that, when this happens, the material in the stream behaves the way a common envelope would. In other words, if the material is not ejected, then the binary must merge. This appears to be somewhat counterintuitive, as it is conceivable that infalling material might be accreted directly without becoming part of a common envelope. However, the aforementioned hydrodynamical simulations show that this is not statistically the case for a dominating portion of the mass in the stream.

\subsubsection{Complete Ejection - No Disk}

Conversely, if all the material is ejected, the subsequent behaviour of the inner binary system will behave similarly to a CE ejection. We adopt the prescription recommended by \citealt{2014MNRAS.438.1909D} for this regime, which we briefly summarise here. 

For the mass that passes through L1 to enter the inner binary's Roche Lobe, its binding energy is
\begin{equation}
E=-\frac{G\left(m_{\rm 1}+m_{\rm 2}\right)\left(m_{\rm 3}-m_{\rm c}\right)}{{\lambda}a_{\rm 1}},
\label{alleject1}
\end{equation}
\noindent where $m_{\rm 3}-m_{\rm c}$ is the total amount of mass that is being ejected, and ${\lambda}$ is a parameter that takes into account the fraction of energy injected by the tertiary, as well as a missing coefficient that describes the relation between $a_{\rm 1}$ and the position from where the material is ejected, which is conspicuously missing from Eq. \ref{alleject1}. In order to eject this mass, this binding energy must be equal to the amount of energy provided by the inner binary:
\begin{equation}
E=\left(\frac{Gm_{\rm 1}m_{\rm 2}}{2a_{\rm 1}}-\frac{Gm_{\rm 1}m_{\rm 2}}{2a_{\rm eject}}\right){\alpha},
\label{alleject2}
\end{equation}
\noindent where ${\alpha}$ is the energy conversion efficiency, and $a_{\rm eject}$ is the post-ejection orbital separation of the inner binary. Needless to say, the inner binary provides this unbinding energy by depleting its own orbital energy. We therefore have
\begin{equation}
a_{\rm eject}=\frac{\mu}{\frac{\mu}{a_{\rm 1}}+\frac{2\left(m_{\rm 3}-m_{\rm c}\right)}{{\alpha}{\lambda}a_{\rm 1}}},
\label{alleject3}
\end{equation}
\noindent where ${\mu}=m_{\rm 1}m_{\rm 2}/\left(m_{\rm 1}+m_{\rm 2}\right)$ is the reduced mass of the inner binary. For our purposes, the greatest uncertainty here is the value of ${\alpha}{\lambda}$, which \cite{2014MNRAS.438.1909D} find to be about 5, but studies on more traditional binary CEs \citep{2013A&A...557A..87T} find to be about 0.3. Due to the uncertainty regarding CEs in general, we prefer not to make any comments on which is the more appropriate value here. Hence, we perform our calculations for both values, and note that the combination of the two values should provide a reasonable lower and upper limit on the effects of the mass ejection.


\subsection{Disk Forms}

If a circumbinary disk is formed around the inner binary after tertiary RLOF, the subsequent dynamical interactions within the inner binary + disk system will not allow the disk to remain in place indefinitely. Thus, again, a certain proportion of the mass in the disk between 0 and 1 is accreted, and the rest must be ejected from the system. In the spirit of our prescription above, we once again posit that the two extremes of complete accretion and complete ejection straddle the parameter space of all possibilities, and that what actually happens must lie somewhere in between.

\subsubsection{Complete Accretion - Disk}

In the case of complete accretion, in which all the infalling material from the tertiary is accreted, we start off with the expression for the initial angular momentum of the inner binary system:
\begin{equation}
J^{\rm 2}=G\frac{m_{\rm 1}^{\rm 2}m_{\rm 2}^{\rm 2}}{m_{\rm 1}+m_{\rm 2}}a_{\rm 1}\left(1-e_{\rm 1}^{\rm 2}\right),
\label{expressJ}
\end{equation}
\noindent which can be manipulated to yield
\begin{equation}
a_{\rm 1}=J^{\rm 2}\frac{m_{\rm 1}+m_{\rm 2}}{Gm_{\rm 1}^{\rm 2}m_{\rm 2}^{\rm 2}}\left(1-e_{\rm 1}^{\rm 2}\right)^{\rm -1},
\label{expressai}
\end{equation}
\noindent likewise, after all the mass has been accreted, we have 
\begin{equation}
a_{\rm accrete}=J_{\rm f}^{\rm 2}\frac{M_{\rm 1}+M_{\rm 2}}{GM_{\rm 1}^{\rm 2}M_{\rm 2}^{\rm 2}}\left(1-e_{\rm 1,f}^{\rm 2}\right)^{\rm -1},
\label{expressaf}
\end{equation}
\noindent where $a_{\rm accrete}$ is the post-accretion inner binary orbital separation, $J_{\rm f}$ is the post-accretion angular momentum, $M_{\rm 1}$ and $M_{\rm 2}$ are the post-accretion values for $m_{\rm 1}$ and $m_{\rm 2}$ respectively, and $e_{\rm 1,f}$ is the post-accretion inner binary eccentricity. For heavily circularised inner orbits, $e_{\rm 1}=e_{\rm 1,f}=1$. Therefore, we have
\begin{equation}
a_{\rm accrete}=a_{\rm 1}\frac{J_{\rm f}^{\rm 2}}{J^{\rm 2}}\frac{m_{\rm 1}^{\rm 2}m_{\rm 2}^{\rm 2}}{m_{\rm 1}+m_{\rm 2}}\frac{M_{\rm 1}+M_{\rm 2}}{M_{\rm 1}^{\rm 2}M_{\rm 2}^{\rm 2}}.
\label{allaccr1}
\end{equation}

In order to find $a_{\rm accrete}$ via Eq. \ref{allaccr1}, we need to find $(M_{\rm 1}+M_{\rm 2})/M_{\rm 1}^{\rm 2}M_{\rm 2}^{\rm 2}$, and $J_{\rm f}^{\rm 2}/J^{\rm 2}$. 

According to \cite{2019ApJ...876L..33P}, when an inner binary accretes mass from a tertiary, it is always the less massive component that receives the lion's share of the mass. We deduce from this that, when the two masses are equal, the accreted mass can be approximated to be divided equally between the two. Therefore, noting that $m_{\rm 2}<m_{\rm 1}$, when 
\begin{equation}
m_{\rm 2}+(m_{\rm 3}-m_{\rm c})<m_{\rm 1},
\label{accrete}
\end{equation}
\noindent we say that 
\begin{equation}
\frac{M_{\rm 1}+M_{\rm 2}}{M_{\rm 1}^{\rm 2}M_{\rm 2}^{\rm 2}}=\frac{m_{\rm 1}+m_{\rm 2}+m_{\rm 3}-m_{\rm c}}{[m_{\rm 2}+(m_{\rm 3}-m_{\rm c})]^{\rm 2}m_{\rm 1}^{\rm 2}},
\end{equation}
\noindent or else
\begin{equation}
\frac{M_{\rm 1}+M_{\rm 2}}{M_{\rm 1}^{\rm 2}M_{\rm 2}^{\rm 2}}=\frac{16}{(m_{\rm 1}+m_{\rm 2}+m_{\rm 3}-m_{\rm c})^{\rm 3}}.
\end{equation}

As for the value of $J_{\rm f}^{\rm 2}/J^{\rm 2}$, many different assumptions could be made. Here, we investigate 3 different possibilities: the prescription used by \cite{2020MNRAS.496.1819L}, the assumption that the angular momentum of the triple system is conserved, and the extreme case in which the infalling material carries no angular momentum at all. 


The prescription preferred by \cite{2020MNRAS.496.1819L} is repeated here:
\begin{equation}
v_{\rm orb,3}a_{\rm 2}\left(1-R_{\rm L}\right)=v_{\rm circ}a_{\rm circ},
\label{leigh1}
\end{equation}
\noindent where $v_{\rm orb,3}$ is the orbital velocity of the L1 Lagrange point of the tertiary relative to the inner binary's COM, $v_{\rm circ}$ and $a_{\rm circ}$ are, respectively, the circular orbital velocity and orbital separation of the accreted matter relative to the inner binary's COM. This basically assumes that the specific angular momentum carried by the accreted mass is the same as that of an object that stays at the L1 Lagrange point of the outer orbit, and that the distance from the inner binary's COM to L1 can be approximated as $a_{\rm 2}\left(1-R_{\rm L}\right)$. The details of why this approximation is valid is provided in Appendix \ref{appA}. This prescription can be expressed as 
\begin{equation}
|{\bf J}_{\rm f}-{\bf J}|=(m_{\rm 3}-m_{\rm c})a_{\rm 2}^{\rm 2}\left(1-R_{\rm L}\right)^{\rm 2}{\omega},
\label{leigh2}
\end{equation}
\noindent where ${\omega}$ is the initial angular velocity of the outer orbit, the blackfont is used to denote vectors, and the angle between ${\bf J}_{\rm f}$ and ${\bf J}$ is simply the inclination angle $i$ between the inner and outer orbits. Noting that ${\omega}^{\rm 2}a_{\rm 2}^{\rm 3}=G(m_{\rm 1}+m_{\rm 2}+m_{\rm 3})$, this can be further simplified to 
\begin{equation}
\begin{split}
|{\bf J}_{\rm f}-{\bf J}|=&(m_{\rm 3}-m_{\rm c})\left(1-R_{\rm L}\right)^{\rm 2}\\
&\left[Ga_{\rm 2}(m_{\rm 1}+m_{\rm 2}+m_{\rm 3})\right]^{\frac{1}{2}}.
\end{split}
\label{leigh3}
\end{equation}
\noindent It should also be noted that, strictly speaking, Eq. \ref{leigh3} is an approximation of a differential equation (see Appendix \ref{appA}), under the assumption that the amount of transferred mass is small. The final angular momentum can therefore be calculated via
\begin{equation}
J^{\rm 2}_{\rm f}=J^{\rm 2}+|{\bf J}_{\rm f}-{\bf J}|^{\rm 2}-2J|{\bf J}_{\rm f}-{\bf J}|{\cos}i,
\label{leigh4}
\end{equation}
\noindent where, again, $i$ is the inclination angle between the inner and outer orbits. 


If the angular momentum of the entire system is taken to be conserved, then the specific angular momentum carried by the accreted material must be equal to that of the tertiary. This angular momentum would be divided between the inner and outer orbits, but for our purposes, we only consider the extreme case where it is all dumped onto the inner orbit. Thus, in this alternate case, Eq. \ref{leigh3} ought to be replaced by
\begin{equation}
\begin{split}
|{\bf J}_{\rm f}-{\bf J}|=&\left(m_{\rm 3}-m_{\rm c}\right)\left(\frac{m_{\rm 1}+m_{\rm 2}}{m_{\rm 1}+m_{\rm 2}+m_{\rm 3}}\right)^{\rm 2}\\
&\left[Ga_{\rm 2}(m_{\rm 1}+m_{\rm 2}+m_{\rm 3})\right]^{\frac{1}{2}},
\end{split}
\label{cons1}
\end{equation}
\noindent where, again, the same approximation of a differential equation is applied.

Lastly, we consider the assumption that the infalling material carries no angular momentum at all. In this case, Eq. \ref{leigh3} ought to be replaced by
\begin{equation}
|{\bf J}_{\rm f}-{\bf J}|=0,
\label{zero1}
\end{equation}
\noindent and $J_{\rm f}^{\rm 2}/J^{\rm 2}=1$. 

Here, it should be pointed out that the final orbital plane of the inner binary may not be the same as that prior to the accretion process. However, this does not affect our study. 

\subsubsection{Complete Ejection - Disk}

In the case in which all the material is ejected from the system after a disk is formed, the material is either ejected directly before it is accreted, in which case a CE is formed and Eq. \ref{alleject3} applies, or it is accreted first before it is re-ejected. In the latter case, we first calculate the angular momentum $J_{\rm f}$ via either Eq. \ref{leigh3}, \ref{cons1}, or \ref{zero1}, and calculate the post-re-ejection inner binary semimajor axis $a_{\rm ej}$ by assuming that the accreted material all leaves the system, carrying the same specific angular momentum as the less massive of the inner binary components after the accretion:
\begin{equation}
a_{\rm ej}=J_{\rm ej}^{\rm 2}\frac{m_{\rm 1}+m_{\rm 2}}{Gm_{\rm 1}^{\rm 2}m_{\rm 2}^{\rm 2}},
\label{reej1}
\end{equation}
\noindent where it is already assumed that the orbit is circularised, and the expression for the post-re-ejection inner binary angular momentum $J_{\rm ej}$ is
\begin{equation}
J_{\rm ej}=J_{\rm f}\left[1-\frac{m_{\rm 3}-m_{\rm c}}{M_{\rm 2}}\frac{M_{\rm 1}}{M_{\rm 1}+M_{\rm 2}}\right],
\label{reej2}
\end{equation}
\noindent where it should be noted that $M_{\rm 2}{\leq}M_{\rm 1}$, and $M_{\rm 1}+M_{\rm 2}=m_{\rm 1}+m_{\rm 2}+m_{\rm 3}-m_{\rm c}$. When Eq. \ref{accrete} is satisfied, 
\begin{equation}
M_{\rm 1}=m_{\rm 1}, 
\label{reej3}
\end{equation}
\noindent and when Eq. \ref{accrete} is not satisfied, 
\begin{equation}
M_{\rm 1}=(m_{\rm 1}+m_{\rm 2}+m_{\rm 3}-m_{\rm c})/2.
\label{reej4}
\end{equation}

It may be argued that it is possible that some, not all, of the material is accreted prior to ejection. However, in this case the material that undergoes accretion must be some proportion between 0 and 1, and the result hence lies in between the two cases mentioned above.

\subsection{Summary of Formation Channel Prescriptions}

Thus, for any system, whether or not a disk forms, and regardless of how much mass is accreted, ejected, or accreted and re-ejected, we can provide a set of values for the final inner binary orbit for each extreme scenario. 

\begin{figure*}
\includegraphics[scale=0.4, angle=0, trim= 0cm 0cm 0cm 0cm]{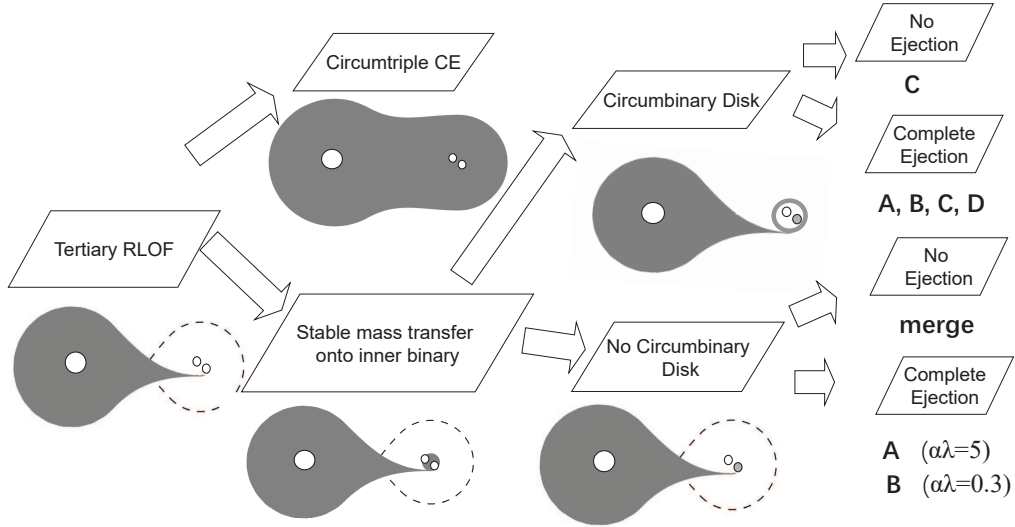}
\caption{Flowchart representation of some possibilities of how a hierarchical triple system can evolve following Roche Lobe overflow from the tertiary. Some individual evolutionary regimes can be characterised by one or more of the 4 extreme scenarios we study, Scenarios A, B, C, and D, and are labelled correspondingly. The final 4 possibility regimes in the right-hand side of the diagram correspond to extreme cases, and what actually happens lies somewhere in between. \label{cartoon}}
\end{figure*}

In summary, other than the regime in which no disk is formed and all the material is accreted, in which case a merger of the inner binary is unavoidable, a total of 4 extreme scenarios need to be considered. These 4 extreme scenarios are as follows: complete ejection without accretion and ${\alpha}{\lambda}=5$, complete ejection without accretion and ${\alpha}{\lambda}=0.3$, complete accretion from a disk, and disk accretion with complete re-ejection following accretion. It should be noted here that these scenarios do not correspond to the 4 different evolutionary regimes induced by the disk / no disk and complete ejection / accretion dichotomies, and we purposefully refer to one by the term ``scenarios" and the other ``regimes" to avoid confusion. Once the results under these 4 extreme scenarios have been obtained, The final fate of the inner binary must lie somewhere in between these 4 sets of results. We proceed to name these 4 extreme scenarios A, B, C, and D respectively, and a summary of where they apply is provided in Fig. \ref{cartoon}.

\section{Data \& Analysis}

\subsection{Our Sample}
To find a theoretical population of triples that is capable of being barium star progenitor candidates, we look to the recent studies conducted by \cite{2020A&A...640A..16T}. In their study, they performed a population synthesis study of hierarchical triples, in which the initial parameters are determined as follows. 

The masses of the more massive of the inner binary components are drawn from a Kroupa IMF \citep{1993MNRAS.262..545K}, while the remaining masses are determined either by sampling a uniform distribution or a distribution used by \cite{2009MNRAS.399.1471E} for $q_{\rm 1}$ and $q_{\rm 2}$. The orbital separations $a_{\rm 1}$ and $a_{\rm 2}$ are assumed to be either uniform in $\log{a}$, log-normal, or follow the distribution used by \cite{2009MNRAS.399.1471E}. The eccentricities of all orbits are assumed to follow a thermal distribution, and the inclinations between the inner and outer orbits are taken to be uniform in $\cos{i}$. A total of ${\sim}60$ thousand systems are generated in this way. Due to the differences in the assumptions regarding the underlying distributions of $q_{\rm 1}$, $q_{\rm 2}$, $a_{\rm 1}$, and $a_{\rm 2}$, this sample is not homogeneous, but rather divided into three sub-samples, named ``OBin", ``T14", and ``E09" in \citealt{2020A&A...640A..16T}. However, for our purposes, we find no reason to distinguish between the three, and therefore proceed to combine them into one sample which we treat as a single initial population.

This initial population is then evolved using the triple evolution code TRES \citep{2016ComAC...3....6T}, which is notably one of the few existing codes which take into account stellar evolution and dynamics simultaneously in a consistent way. This results in a sample of evolved systems, which reasonably approximates observed triple populations, albeit with all the biases and selection effects inherent to such observations. 

For our following analysis, we start with the same sample.

\subsection{Preliminary Processing}

From the sample detailed above, we select all hierarchical triple systems where the tertiary has a large enough radius to fill its Roche Lobe during the AGB phase. This leads to a preliminary sample of 440 systems, the tertiaries of which all have masses less than 7.5$M_{\odot}$, and therefore have no danger of undergoing core collapse. This sample of 440 corresponds to a population of stellar systems that have a total birthrate of $4.2{\times}10^{-4}$/M$\odot$ for a given stellar population, or a Galactic rate of $1.3{\times}10^{-3}$/yr \citep{2020A&A...640A..16T}. Of these, we find 43 systems in which their tertiaries can already fill their Roche Lobes during the RGB phase; while these systems are expected to undergo processes similar to those herein studied, and consequently can potentially form Ba stars, the subsequent evolution following RGB RLOF is too complicated to be covered in this paper. Hence, in the interest of the uniformity of our sample, we eliminate them. We also find 5 systems where the components of the inner binary are not MS stars when the tertiary fills its Roche Lobe; but, as we see no reason why these could not form Ba stars, we include them in our sample, and only mention their existence here for the record. 

Of the remaining 397 systems, 243 systems are found to undergo circumtriple CEs after tertiary RLOF commences, according to the criteria outlined in Eq. \ref{outCEcrit}. Eliminating these, we arrive at our final sample of 154 Ba star progenitor candidates. See Table \ref{sample} for a summary of this selection process.

\begin{table*}
	\centering
	\caption{Initial preprocessing of our data sample.}
	\label{sample}
	\begin{tabular}{ccc}
		\hline
		Selection Criteria & \# of systems after selection & proportion to total \# of systems simulated \\
		\hline
                tertiary AGB RLOF & 440 & 0.73\% \\
                no pre-AGB RLOF & 397 & 0.66\% \\
                no circumtriple CE & 154 & 0.26\% \\
		\hline
	\end{tabular}
\end{table*}

It is interesting to note that, for all 154 systems, the combined mass of the inner binary is greater than that of the tertiary. Granted, this is partly due to a selection effect imposed by our sample: when \citealt{2020A&A...640A..16T} generated their sample of hierarchical triple systems which we are using here, the mass ratios of two-thirds of the systems were generated using a distribution that guarantees $q_{\rm 2}<1$. However, of the remaining one-third of the ${\sim}60$ thousand systems, many ought to have $q_{\rm 2}{\geqslant}1$. Therefore, some explanation is required as to why none of them end up in the final sample. Our understanding is that this is due to an instability that has a direct analogue in binary evolution. In binary evolution, systems undergoing conservative mass transfer from a more massive to a less massive object experience greater mutual gravitational attraction between their components with the passing of time, and hence tend to see their orbits shrink. This creates a positive feedback loop in which mass tranfer accelerates with shrinking orbital separation, and orbital separations shrink faster with increased mass transfer rates, ultimately rendering the mass transfer unstable. Similarly, systems in which $m_{\rm 1}+m_{\rm 2}<m_{\rm 3}$ will undergo a similar feedback loop if $m_{\rm 3}$ were to initiate any sort of mass transfer prior to becoming an AGB star, thus eliminating many such systems from our sample. The lack of these systems reflects the difficulty of forming Ba stars through mass transfer from a more massive star to a less massive star in a binary system, which is the very same mechanism that causes many binary Ba star progenitors to fail.


We then proceed to determine how many of these 154 systems give birth to Ba stars. Since these systems are all undergoing AGB RLOF from their tertiaries, they are accreting Ba-rich material. Since only a small amount of Ba, equivalent to only 2-30 times what is expected in a regular MS star, is required to turn such accretors into Ba stars, it is highly probable that the inner binary components will have enough combined Ba to identify as a Ba star at the end of the accretion. The only issue is whether they will merge.


\subsection{Effects of Tertiary Tides}

Before the tertiary undergoes RLOF, its radius will have become great enough to allow significant tertiary tides (TTs) to come into effect (see \citealt{2018MNRAS.479.3604G}). This would decrease the inner binary orbital separation, increasing the possibility that the inner binary would merge, thereby potentially leading to an increased Ba star formation rate. It should be noted that the simulations conducted by \cite{2020A&A...640A..16T}, from which we obtain our initial sample, do not take this effect into account. 

\subsubsection{Prescription for Tertiary Tidal Influence}

Physically speaking, for a system in our sample, TTs are significant during two phases of its evolution: once during the tertiary's RGB phase, and once during its AGB phase, as the mass transfer is occurring. This is because the radius of the tertiary is large during these two phases, which is a prerequisite for decreasing the inner binary separation via this mechanism. To account for this effect, we apply the results of \cite{2020MNRAS.491..264G}, repeated here:
\begin{equation}
\begin{split}
\frac{ 1  }{  a_{\rm 1}   } \frac{ {\rm d}a_{\rm 1} } { {\rm d}t }=&\left(2.22{\times}10^{-8}{\rm yrs}^{-1}\right)    \frac{4q}{\left(1+q\right)^{2}}\left(\frac{R_{\rm 3}}{100{\rm R}_{\odot}}\right)^{5.2} \\
&\left(\frac{a_{\rm 1}}{0.2{\rm AU}}\right)^{4.8}\left(\frac{a_{\rm 2}}{2{\rm AU}}\right)^{-10.2}\left(\frac{\tau}{0.534{\rm yrs}}\right)^{-1.0}.
\label{TTresult} 
\end{split}
\end{equation}
to our sample of 154 Ba star progenitor candidates. To do this, we simplify Eq. \ref{TTresult} into 
\begin{equation}
\begin{cases}
&a_{\rm f}=\left(4.8Ct+a_{\rm 1}^{-4.8}\right)^{-\frac{1}{4.8}},\\
&C{\equiv}C\left(q,R_{\rm 3},a_{\rm 2},{\tau}\right)=\\
&\left(2.22{\times}10^{-8}{\rm yrs}^{-1}\right)\frac{4q}{\left(1+q\right)^{2}}\left(\frac{R_{\rm 3}}{100{\rm R}_{\odot}}\right)^{5.2}\\
&\left(\frac{1}{0.2{\rm AU}}\right)^{4.8}\left(\frac{a_{\rm 2}}{2{\rm AU}}\right)^{-10.2}\left(\frac{\tau}{0.534{\rm yrs}}\right)^{-1.0}.
\end{cases}
\label{TTsimp}
\end{equation}
\noindent where $q=m_{\rm 2}/m_{\rm 1}$, $R_{\rm 3}$ is the tertiary radius, $t$ is the time over which TTs take effect, and $\tau$ is the viscoelastic relaxation time. Of these parameters, the value of $\tau$ for a given system is the most poorly understood, and will be the primary source of uncertainty when we seek to estimate the magnitude of TT effects on each system.

Tertiary tidal effects are comparitively well studied for the period of the RGB phase in which the tertiary is expanding (in which $R_{\rm 3}$ increases in size), during which the value of $\tau$ has been determined to be about 0.019 years for HD97131, although typical values for other systems usually appear to be longer \citep{2018MNRAS.479.3604G}. For the period of the RGB phase during which the tertiary's radius is contracting, it is relatively poorly understood, but studies of HD1810168 suggest that the value of $\tau$ decreases drastically during this period due to resonant locking \citep{2013MNRAS.429.2425F}. No similar study has yet been attempted for what happens during the tertiary's AGB phase, but similarities in the structure of the tertiary between the RGB and AGB phases indicate that the process should be analogous.

\subsubsection{Parameter Selection}

Needless to say, in order to make the simplification of Eq. \ref{TTsimp}, $q$, $R_{\rm 3}$, $a_{\rm 1}$, $a_{\rm 2}$, and $\tau$ all need to be constant. It should also be known that Eq. \ref{TTresult} was obtained under the assumption of circular and coplanar inner and outer orbits. For simplicity, and considering the aforementioned limitations of our prescription, for all systems in our sample, we model the effects of tertiary tides as a one-off adjustment of $a_{\rm 1}$ due to TTs prior to the tertiary RLOF. When calculating this one-off adjustment, we assume that the values of $q$, $R_{\rm 3}$, $a_{\rm 1}$, and $a_{\rm 2}$ are the same as those found at the onset of AGB RLOF for each system in question, and that the tertiary tidal influence persists over a period of $t=10^6$ years, with a $\tau$ value of $10^{-4}$ years. The value for $t$ is chosen to represent the total amount of time that a star spends close to maximum radius in its RGB and AGB evolution phases combined, and the value of $\tau$ is chosen to represent a typical viscoelastic relaxation time with resonant locking taken into account. For systems with orbits which are non-coplanar, we assume coplanar orbits for the purpose of this particular calculation, since the effects of deviations from coplanarity in the context of TTs is poorly understood. Likewise, for systems with eccentric orbits, we assume that the orbits are circularised under conservation of angular momentum:
\begin{equation}
a_{\rm i,circ}=a_{\rm i}\left(1-e_{\rm i}^{\rm 2}\right),
\label{orbcirc}
\end{equation}
\noindent where $i=1,2$. This should provide an estimate of the effects of TTs on the subsequent evolution of the systems, after we replace the values of $a_{\rm 1}$ with $a_{\rm f}$ in Eq. \ref{TTsimp}. Due to the uncertainties regarding the exact magnitude of TT effects, we also perform the same subsequent calculations using the original value of $a_{\rm 1}$, without accounting for TTs, for comparison.

\subsubsection{Accounting for Tertiary Tides}

\begin{figure}
\includegraphics[scale=0.25, angle=0, trim= 2cm 0cm 0cm 0cm]{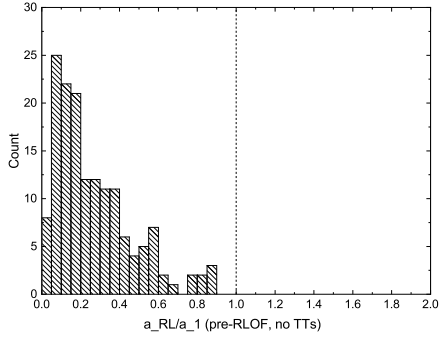}
\caption{Initial distribution of $a_{\rm RL}/a_{\rm 1}$ for our sample immediately prior to tertiary RLOF, without accounting for TTs. The dashed line represents the value at which the inner binary undergoes mass transfer via Roche Lobe overflow. \label{fig2}}
\end{figure}

\begin{figure}
\includegraphics[scale=0.25, angle=0, trim= 2cm 0cm 0cm 0cm]{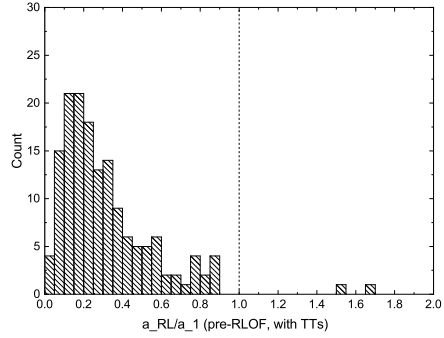}
\caption{Initial distribution of $a_{\rm RL}/a_{\rm 1}$ for our sample immediately prior to tertiary RLOF, having accounted for TTs. The dashed line represents the value at which the inner binary undergoes mass transfer via Roche Lobe overflow. It can be seen that 2 systems are already have inner binaries interacting via mass transfer. \label{fig3}}
\end{figure}

The distribution of the initial values of $1/a_{\rm 1}$ with and without accounting for TTs at the onset of tertiary AGB RLOF are shown in Figs. \ref{fig2} and \ref{fig3}. This is expressed as $a_{\rm RL}/a_{\rm 1}$, where $a_{\rm RL}$ is the value required to start the inner binary RLOF. The advantage of plotting our data in this way lies in the ease of determining whether or not a RLOF has occurred in the inner binary- if this value is 1 or greater, then inner binary RLOF commences for the system in question.

In all but two cases, TTs alone are not sufficient to drive the inner binary to interact prior to the RLOF of the tertiary. For our purposes, we treat RLOF within the inner binary as a proxy for the two binary components being close enough to merge. As such, we assume that these two systems form Ba stars. The same proxy is likewise used in our following analysis.

Here, it should be noted that changing the inner binary separation will not affect whether or not the tertiary can fill its Roche Lobe, and all other orbital parameters of the hierarchical triple system remain practically the same. It may influence whether or not a circumbinary disk forms around the inner binary following AGB RLOF of the tertiary, and hence may affect whether this RLOF results in a merging event within the inner binary. However, given how agnostic our treatment is regarding whether or not such a disk forms, it does not influence our following analysis. 

\subsection{Post-RLOF Inner Binary Calculations}

For those hierarchical triple systems that survive the effects of TTs, as well as those in the sample for which we neglect the effects of TTs altogether, there is no known effect which would henceforth prevent their tertiaries from reaching the AGB phase, and filling their Roche Lobes. Thereafter, these systems would undergo the processes which we have discussed in the previous section, and their effects on the inner binary can be calculated accordingly.

For simplicity, we assume that all orbits circularise rapidly according to Eq. \ref{orbcirc} following the onset of RLOF. Each system may or may not form a circumbinary disk, but due to the uncertainty regarding whether this happens or not, we do not make this distinction for each individual system. We instead proceed to calculate the final distributions of $a_{\rm RL}/a_{\rm 1}$ under the assumption that all systems in our sample uniformly evolve through one of each of the previously mentioned scenarios, A, B, C, and D. As previously mentioned, these scenarios correspond to no disk with complete ejection without accretion and ${\alpha}{\lambda}=5$, no disk with complete ejection without accretion and ${\alpha}{\lambda}=0.3$, complete accretion with or without a disk, and disk with complete re-ejection following accretion, respectively. This yields 4 different distributions for the 4 different scenarios. It should be pointed out that, aside from the possibility that some of these systems will simply fail both at forming a disk and ejecting the subsequent common envelope, the true evolutionary result would lie somewhere in between the 4 different scenarios, as each of the 4 scenarios address a different extreme assumption of inner binary.

For Scenario A evolution, we apply Eq. \ref{alleject3}, under the assumption that ${\alpha}{\lambda}=5$, and take $a_{\rm eject}$ to be the final value for $a_{\rm 1}$.

For Scenario B evolution, we apply the same equation as Scenario A, under the assumption that ${\alpha}{\lambda}=0.3$, and again take $a_{\rm eject}$ to be the final value for $a_{\rm 1}$.

For Scenario C evolution, we adopt Eq. \ref{allaccr1}. For the case when we adopt the treatment of \cite{2020MNRAS.496.1819L}, the relevant variables are determined by Eqs. \ref{accrete} to \ref{leigh4}. For the case where we assume that the angular momentum of the triple system is conserved, we substitute Eq. \ref{leigh3} with Eq. \ref{cons1}. For the extreme case in which the accreted material does not have any angular momentum, we substitute Eq. \ref{leigh3} with Eq. \ref{zero1}. Finally, in each case, $a_{\rm accrete}$ is used for the final value for $a_{\rm 1}$. 

For Scenario D evolution, we first calculate $a_{\rm accrete}$ according to Scenario C evolution, but afterwards apply Eqs. \ref{reej1} to \ref{reej4}, and take $a_{\rm ej}$ to be the final value for $a_{\rm 1}$.

It should be noted that, for most of the treatments above, they are only valid for circular orbits, which we have already ascertained by circularising all the orbits via Eq. \ref{orbcirc}.

\subsection{Outer Orbital Periods}

For the Ba stars originating from the hierarchical triple formation channel we consider, if they are to account for the observed Ba star binaries which lie in the theoretical period gap at 1000 days, the final periods of the outer orbits must fall within this gap. To test for this possibility, we treat the hierarchical triple system as a binary with two bodies of mass $m_{\rm 1}+m_{\rm 2}$ and $m_{\rm 3}$ respectively, and proceed to investigate whether this system can have a final orbital period within this period gap. This approach will be attempted in the next section, while here we lay down some of the equations that are relevant for such an undertaking.

Assuming that only a fraction $\beta$ (which is constant with time) of the mass transferred from the tertiary is retained by the inner binary, and assuming isotropic re-emission for the mass lost from the system, the evolution of the outer orbit will follow
\begin{equation}
\begin{cases}
\frac{\dot{a}_{\rm 2}}{a_{\rm 2}}&=-2\frac{\dot{m}_{\rm 3}}{m_{\rm 3}}\left[1-{\beta}\frac{m_{\rm 3}}{m_{\rm 1}+m_{\rm 2}}-\right.\\
&\left.\left(1-{\beta}\right)\left(\frac{m_{\rm 3}}{m_{\rm 1}+m_{\rm 2}}+\frac{1}{2}\right)\frac{m_{\rm 3}}{m_{\rm 1}+m_{\rm 2}+m_{\rm 3}}\right],\\
\left(\frac{P_{\rm final}}{P_{\rm init}}\right)^{\rm 2}&=\left(\frac{a_{\rm 2,final}}{a_{\rm 2,init}}\right)^{\rm 3}\\
&\frac{m_{\rm 1}+m_{\rm 2}+m_{\rm 3}}{m_{\rm 1}+m_{\rm 2}+m_{\rm 3}-\left(1-{\beta}\right)\left(m_{\rm 3}-m_{\rm c}\right)},
\end{cases}
\label{outP1}
\end{equation}
\noindent where $P_{\rm init}$ and $P_{\rm final}$ are the initial and final outer orbital periods respectively, while $a_{\rm 2,init}$ and $a_{\rm 2,final}$ are the initial and final semi-major axes respectively. The first equation of \ref{outP1} can be manipulated to yield
\begin{equation}
\begin{split}
\frac{\dot{a}_{\rm 2}}{a_{\rm 2}}=&-2\frac{\dot{m}_{\rm 3}}{m_{\rm 3}}+2{\beta}\frac{\dot{m}_{\rm 3}}{m_{\rm 1}+m_{\rm 2}}+\\
&2\left(1-{\beta}\right)\frac{\dot{m}_{\rm 3}}{m_{\rm 1}+m_{\rm 2}}\frac{m_{\rm 3}}{m_{\rm 1}+m_{\rm 2}+m_{\rm 3}}+\\
&\left(1-{\beta}\right)\frac{\dot{m}_{\rm 3}}{m_{\rm 1}+m_{\rm 2}+m_{\rm 3}}\\
=&-2\frac{\dot{m}_{\rm 3}}{m_{\rm 3}}+2\frac{\dot{m}_{\rm 3}}{m_{\rm 1}+m_{\rm 2}}-\\
&\left(1-{\beta}\right)\frac{\dot{m}_{\rm 3}}{m_{\rm 1}+m_{\rm 2}+m_{\rm 3}}.
\end{split}
\label{outP2}
\end{equation}
\noindent For conservative mass transfer (${\beta}=1$), noting that $d\left(m_{\rm 1}+m_{\rm 2}\right)/dt=-{\beta}\dot{m}_{\rm 3}$, this simplifies to:
\begin{equation}
\begin{cases}
P_{\rm init}^{\rm 2}&=\frac{\left(2{\pi}\right)^{\rm 2}a_{\rm 2}^{\rm 3}}{G\left(m_{\rm 1}+m_{\rm 2}+m_{\rm 3}\right)},\\
\frac{P_{\rm final}}{P_{\rm init}}&=\left(\frac{m_{\rm 3}}{m_{\rm c}}\frac{m_{\rm 1}+m_{\rm 2}}{m_{\rm 1}+m_{\rm 2}+m_{\rm 3}-m_{\rm c}}\right)^{\rm 3}.
\end{cases}
\label{outP3}
\end{equation}
\noindent Similarly, for total mass loss (${\beta}=0$),
\begin{equation}
\begin{cases}
P_{\rm init}^{\rm 2}=&\frac{\left(2{\pi}\right)^{\rm 2}a_{\rm 2}^{\rm 3}}{G\left(m_{\rm 1}+m_{\rm 2}+m_{\rm 3}\right)},\\
\frac{P_{\rm final}}{P_{\rm init}}=&\left(\frac{m_{\rm 3}}{m_{\rm c}}\right)^{\rm 3}{\exp}\left[3\left(\frac{m_{\rm c}-m_{\rm 3}}{m_{\rm 1}+m_{\rm 2}}\right)\right]\left(\frac{m_{\rm 1}+m_{\rm 2}+m_{\rm 3}}{m_{\rm 1}+m_{\rm 2}+m_{\rm c}}\right)^{\rm 2}.
\end{cases}
\label{outP4}
\end{equation}
\noindent where $a_{\rm 2,final}$ is the post-mass-transfer value of $a_{\rm 2}$, and is cancelled out between the second and third equations of Eq. \ref{outP4}.

For everything in between conservative mass transfer and total mass loss ($0<{\beta}<1$), we have 
\begin{equation}
\begin{cases}
P_{\rm init}^{\rm 2}=&\frac{\left(2{\pi}\right)^{\rm 2}a_{\rm 2}^{\rm 3}}{G\left(m_{\rm 1}+m_{\rm 2}+m_{\rm 3}\right)},\\
\frac{P_{\rm final}}{P_{\rm init}}=&\left(\frac{m_{\rm 3}}{m_{\rm c}}\right)^{\rm 3}\left(\frac{m_{\rm 1}+m_{\rm 2}}{m_{\rm 1}+m_{\rm 2}+{\beta}(m_{\rm 3}-m_{\rm c})}\right)^{\rm \frac{3}{\beta}}\\
&\left(\frac{m_{\rm 1}+m_{\rm 2}+m_{\rm 3}}{m_{\rm 1}+m_{\rm 2}+{\beta}(m_{\rm 3}-m_{\rm c})+m_{\rm c}}\right)^{\rm 2}.
\end{cases}
\label{outP5}
\end{equation}
\noindent which is a continuous function of ${\beta}$, and matches the solutions of Eqs. \ref{outP3} and \ref{outP4} at the limits of ${\beta}{\rightarrow}1$ and ${\beta}{\rightarrow}0$, respectively.

Thus, using Eqs. \ref{outP3}, \ref{outP4}, and \ref{outP5}, a final orbital period can be derived for the outer orbit. This can be done for any ${\beta}$, without having to calculate for any of the specifics of the mass transfer process.

\section{Results}

\subsection{Scenario A \& B Evolution}

\begin{figure}
\begin{tabular}{c}
\includegraphics[scale=0.25, angle=0, trim= 2cm 0cm 0cm 0cm]{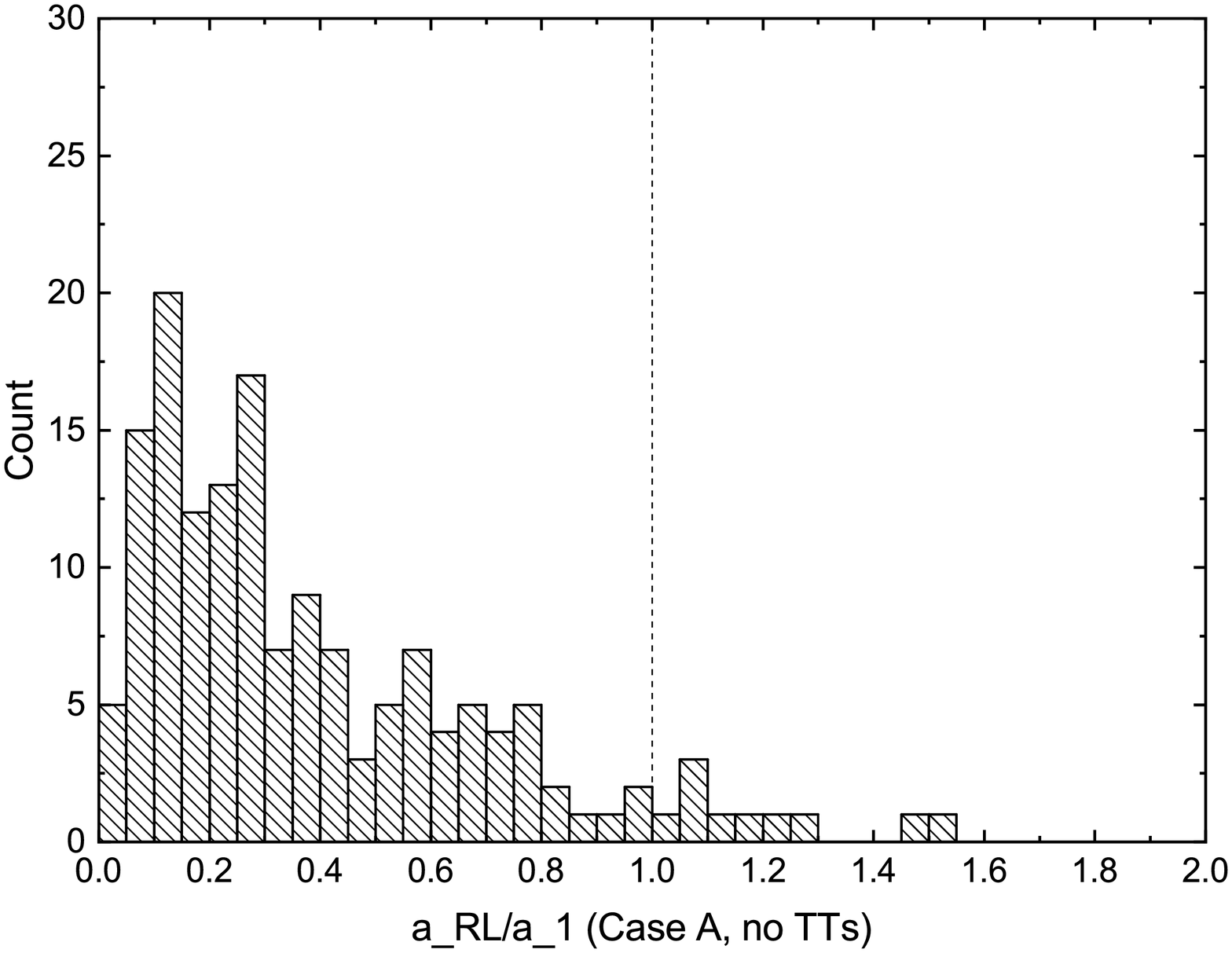}\\
\includegraphics[scale=0.25, angle=0, trim= 2cm 0cm 0cm 0cm]{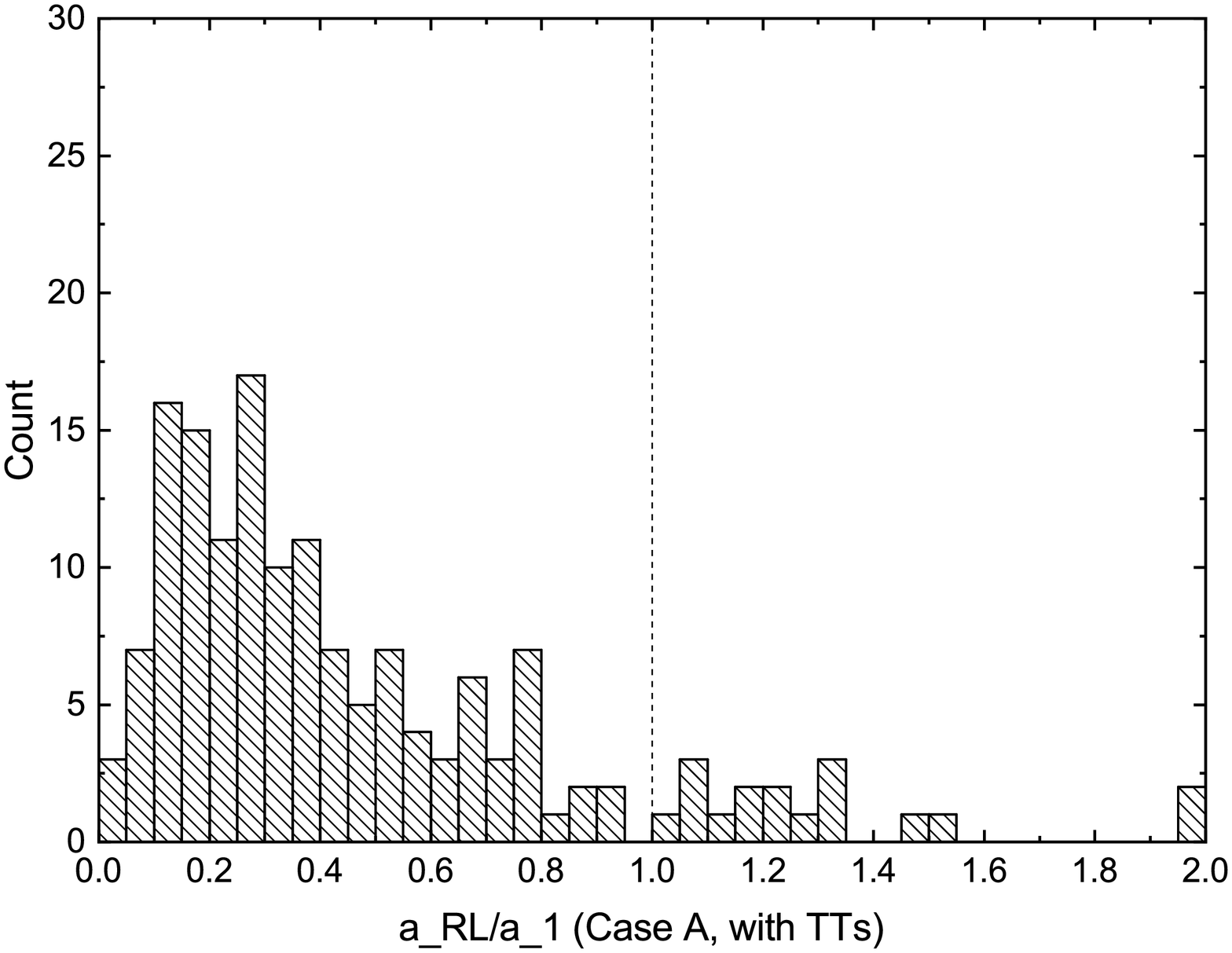}
\end{tabular}
\caption{Results of calculations under the assumption that the mass overflowing from the tertiary undergoes complete ejection without being accreted, and ${\alpha}{\lambda}=5$ (Scenario A). Upper panel: results when tertiary tides are not taken into account. Lower panel: results when tertiary tides have been considered. A value of $a_{\rm RL}/a_{\rm 1}$ greater than 1 (dashed line) indicates that the inner binary is close enough to undergo Roche Lobe Overflow; 10 and 17 systems are undergoing inner binary RLOF in the upper and lower panels, respectively. For aesthetic reasons, all values of $a_{\rm RL}/a_{\rm 1}$ greater than 2 have been binned in the rightmost bin of the plot. \label{scenarioA}}
\end{figure}

\begin{figure}
\begin{tabular}{c}
\includegraphics[scale=0.25, angle=0, trim= 2cm 0cm 0cm 0cm]{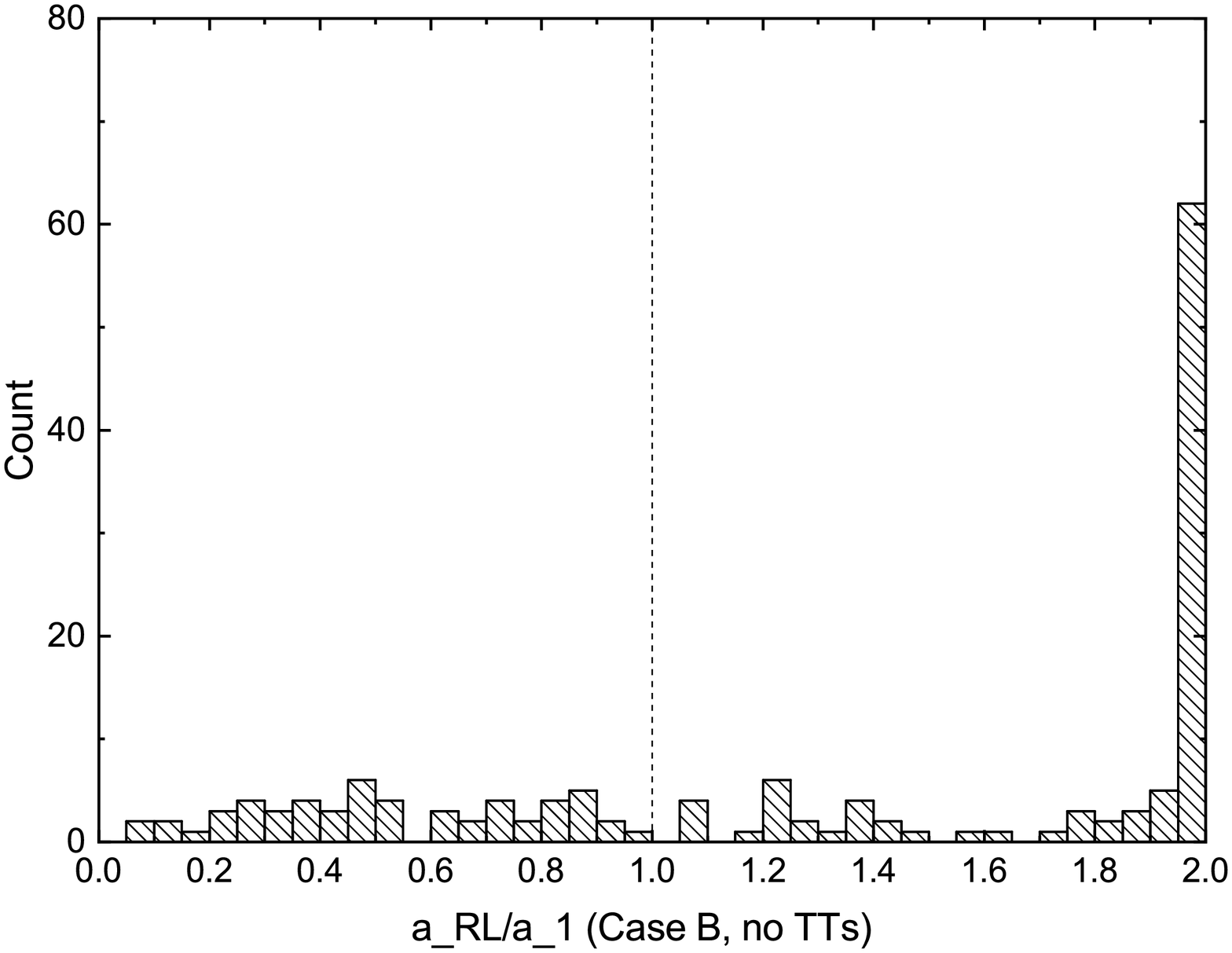}\\
\includegraphics[scale=0.25, angle=0, trim= 2cm 0cm 0cm 0cm]{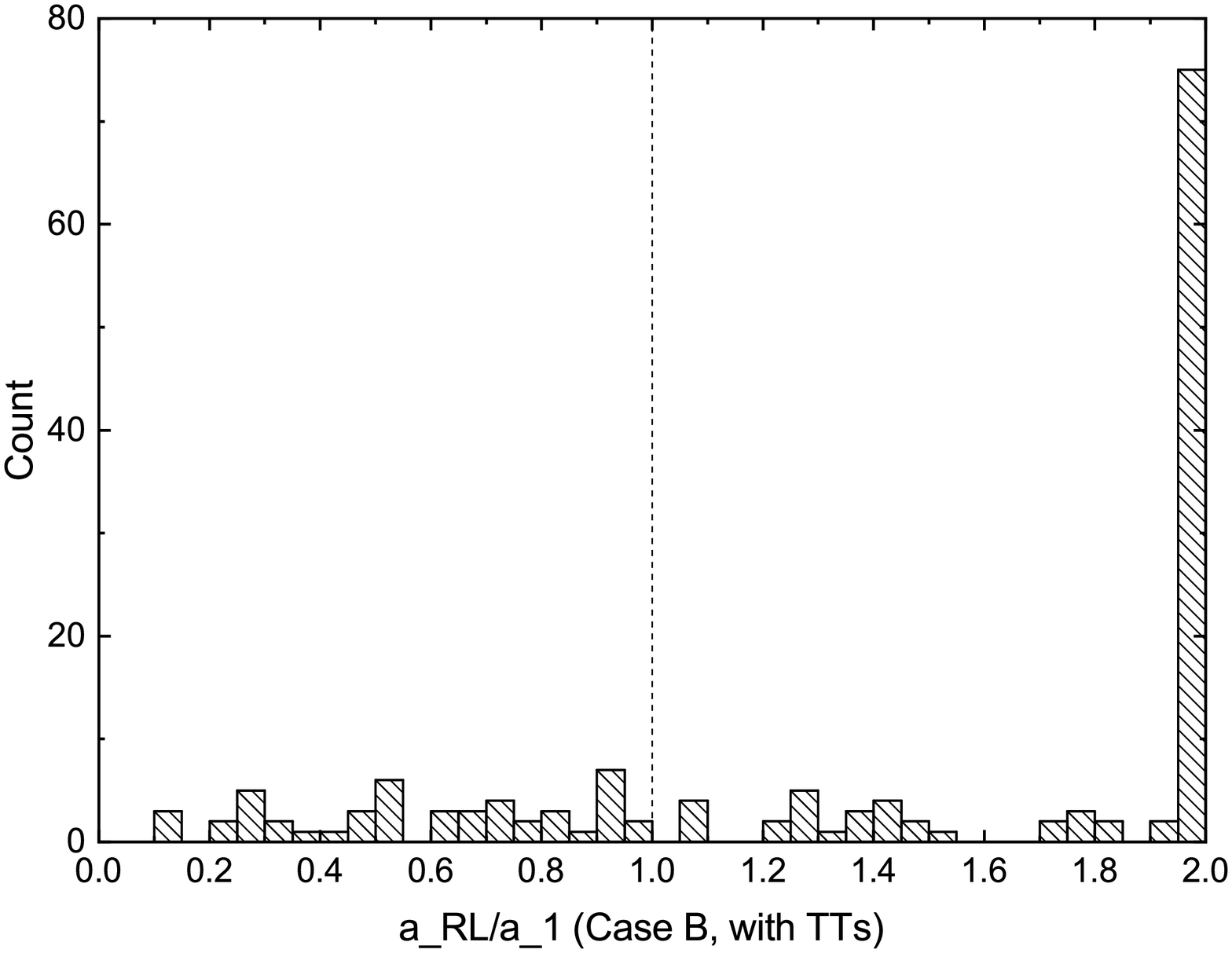}
\end{tabular}
\caption{Results of calculations under the assumption that the mass overflowing from the tertiary undergoes complete ejection without being accreted, and ${\alpha}{\lambda}=0.3$ (Scenario B). Upper panel: results when tertiary tides are not taken into account. Lower panel: results when tertiary tides have been considered. A value of $a_{\rm RL}/a_{\rm 1}$ greater than 1 (dashed line) indicates that the inner binary is close enough to undergo Roche Lobe Overflow; 99 and 106 systems are undergoing inner binary RLOF in the upper and lower panels, respectively. For aesthetic reasons, all values of $a_{\rm RL}/a_{\rm 1}$ greater than 2 have been binned in the rightmost bin of the plot. \label{scenarioB}}
\end{figure}

For the scenario where there is no disk, and a complete ejection of the material inflowing from the tertiary is achieved, the evolution of the triple system corresponds to Scenarios A and B in our taxonomy. The final distributions of $a_{\rm RL}/a_{\rm 1}$ for Scenario A (${\alpha}{\lambda}=5$) and Scenario B (${\alpha}{\lambda}=0.3$) evolution are presented in Figs. \ref{scenarioA} and \ref{scenarioB}, respectively.

For Scenario A evolution, 10 triple systems would be undergoing inner binary RLOF after ejecting the infalling material from the tertiary, and this number increases to 17 if the reduced inner binary separations due to tertiary tides are considered instead of the initial values. Since our original sample of 440 corresponded to a set of hierarchical triple systems with a Galactic rate of $1.3{\times}10^{-3}$/yr \citep{2020A&A...640A..16T}, 10 and 17 triple systems should correspond to a Galactic rate of $3.0{\times}10^{-5}$/yr and $5.0{\times}10^{-5}$/yr respectively. If inner binary RLOF can be seen as a proxy for a merged inner binary, then this would indicate a corresponding number of Ba stars arising from our sample.

Likewise, for Scenario B evolution, 99 Ba stars are expected from our sample in the absence of TTs, whereas 106 are expected if TTs have been taken into account. This corresponds to a Galactic rate of $2.9{\times}10^{-4}$/yr and $3.1{\times}10^{-4}$/yr respectively. The increased Ba star birthrate is a natural result of a less efficient energy conversion rate.

These results are summarised in Tab. \ref{results}.

\subsection{Scenario C Evolution}

\begin{figure}
\begin{tabular}{c}
\includegraphics[scale=0.25, angle=0, trim= 2cm 0cm 0cm 0cm]{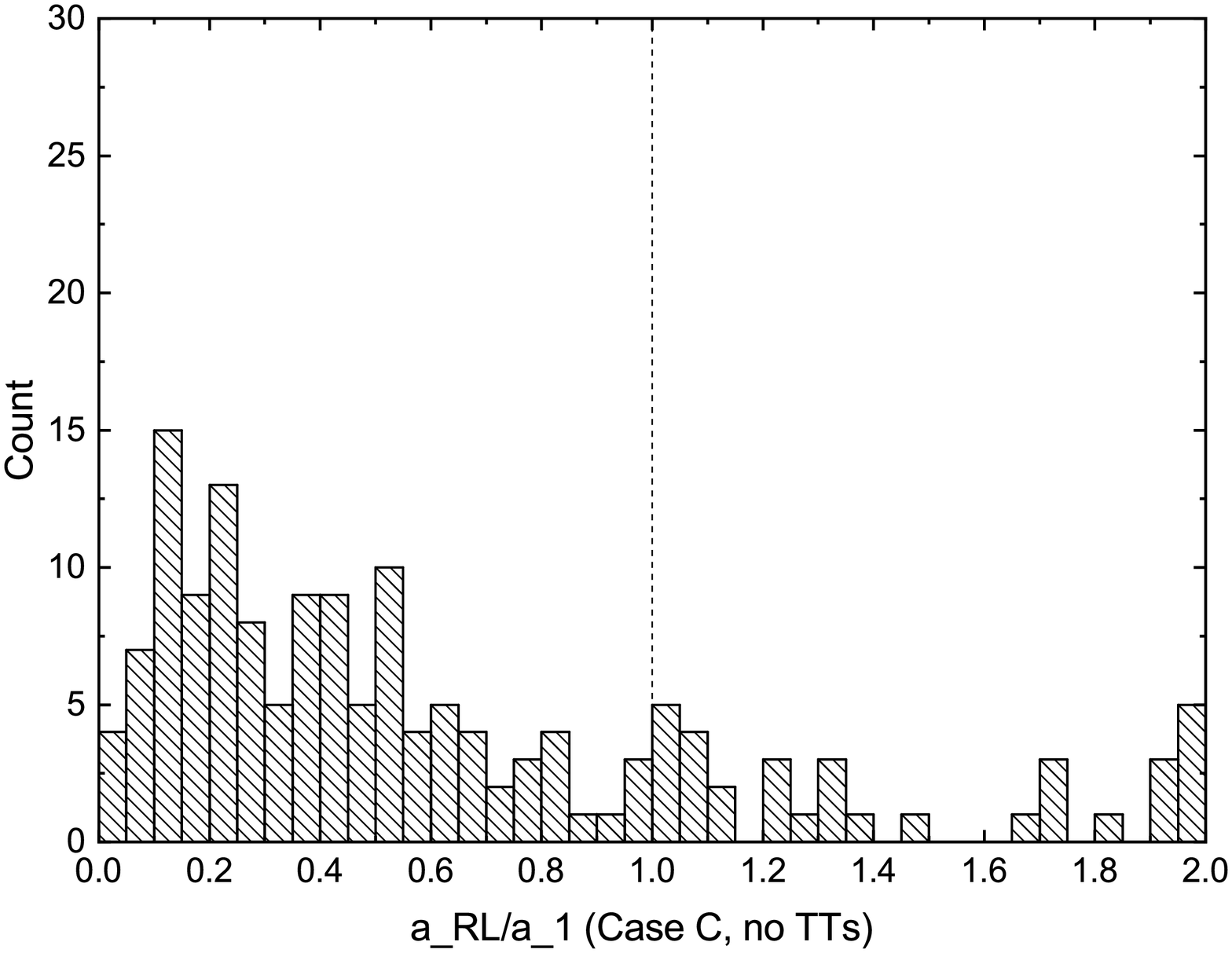}\\
\includegraphics[scale=0.25, angle=0, trim= 2cm 0cm 0cm 0cm]{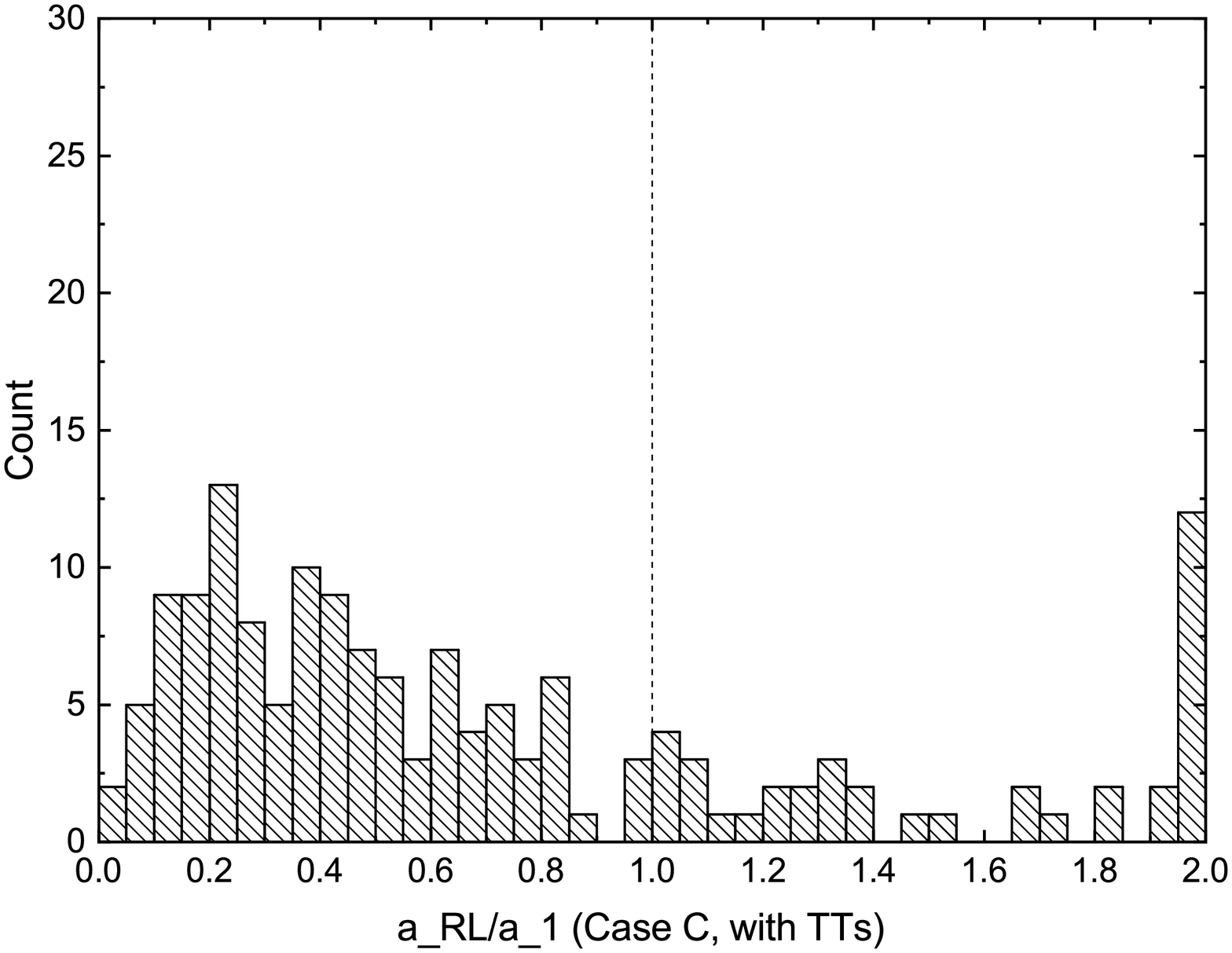}
\end{tabular}
\caption{Results of calculations under the assumption that the mass overflowing from the tertiary is completely absorbed via disk accretion without mass loss (Scenario C), and that the accreted material does not carry any angular momentum. Upper panel: results when tertiary tides are not taken into account. Lower panel: results when tertiary tides have been considered. A value of $a_{\rm RL}/a_{\rm 1}$ greater than 1 (dashed line) indicates that the inner binary is close enough to undergo Roche Lobe Overflow; 33 and 39 systems are undergoing inner binary RLOF in the upper and lower panels, respectively. For aesthetic reasons, all values of $a_{\rm RL}/a_{\rm 1}$ greater than 2 have been binned in the rightmost bin of the plot. \label{scenarioC}}
\end{figure}

\begin{figure}
\begin{tabular}{c}
\includegraphics[scale=0.25, angle=0, trim= 2cm 0cm 0cm 0cm]{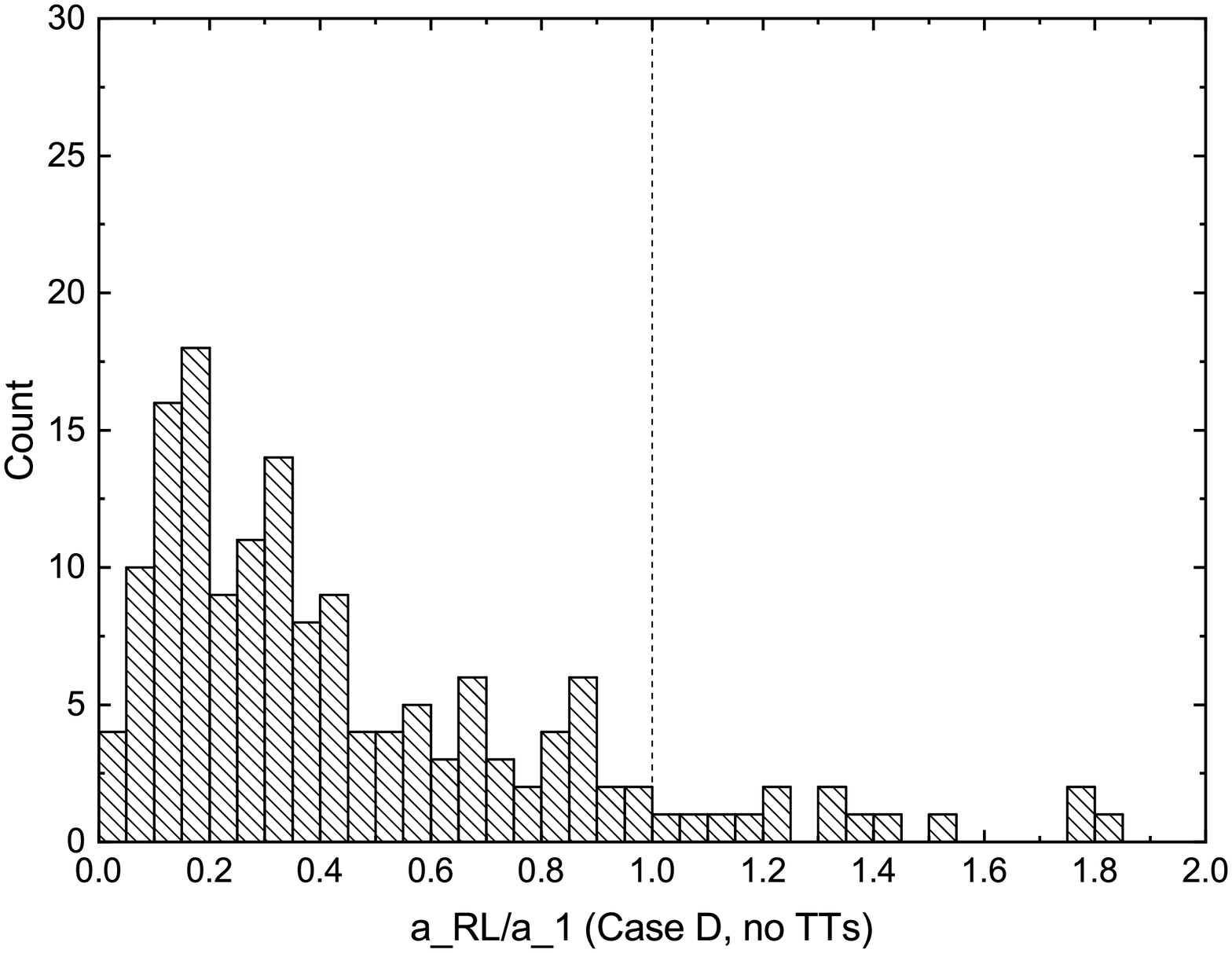}\\
\includegraphics[scale=0.25, angle=0, trim= 2cm 0cm 0cm 0cm]{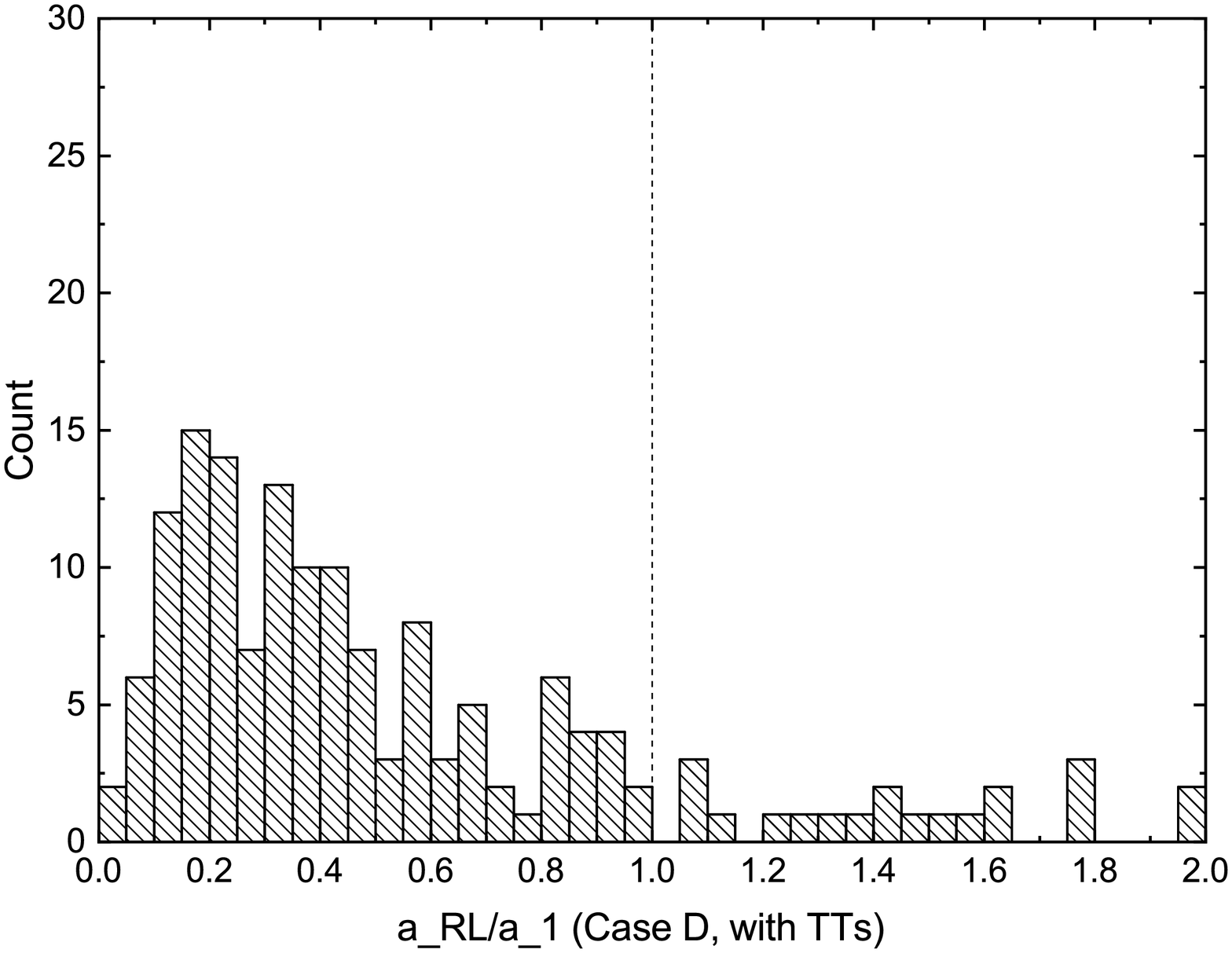}
\end{tabular}
\caption{Results of calculations under the assumption that the mass overflowing from the tertiary undergoes complete ejection after being accreted (Scenario D), and that the accreted material does not carry any angular momentum when accreted, but has the same specific angular momentum as the less massive inner binary component when re-ejected. Upper panel: results when tertiary tides are not taken into account. Lower panel: results when tertiary tides have been considered. A value of $a_{\rm RL}/a_{\rm 1}$ greater than 1 (dashed line) indicates that the inner binary is close enough to undergo Roche Lobe Overflow; 14 and 20 systems are undergoing inner binary RLOF in the upper and lower panels, respectively. For aesthetic reasons, all values of $a_{\rm RL}/a_{\rm 1}$ greater than 2 have been binned in the rightmost bin of the plot. \label{scenarioD}}
\end{figure}

The next scenario we consider is the one in which all the material that is dumped onto the inner binary by the tertiary is absorbed through disk accretion, without any form of mass loss from the inner binary whatsoever. This evolution channel, which we have named Scenario C evolution, covers the other extreme end of possible mass loss ratios when considered in conjunction with the other channels we investigate; when coupled with either Scenario A, B, or D, the result of any accretion scenario must lie in between the 4 scenarios, as the proportion of mass lost in the accretion must lie in between 0 and 1.

As previously explained, the results of this scenario are primarily dependent on how much angular momentum is carried in the material that is accreted. We start by investigating the assumption that angular momentum is conserved in the triple system. This corresponds to Eq. \ref{cons1}. The result of this assumption is that the angular momentum carried by the accreted material is systematically about an order of magnitude greater than that of the inner binary. As such, no inner binaries of any of the 154 hierarchical triple systems merge, and no Ba stars are formed. Granted, inner binaries that are in almost exactly retrograde orbits to their tertiaries may still merge in this scenario, since their initial orbital angular momenta are in the opposite direction of the angular momenta being pumped into it, but it is highly doubtful that the resultant Ba star would be able to survive the extreme spins that would result from the subsequent pumping. In any case, we can be certain that this would result in very few, if any, Ba stars.

Of all the assumptions regarding angular momentum that we attempt for Scenario C, this one can be argued to be the most physical, as it is the only assumption we investigate which guarantees the angular momentum conservation of a closed system. Indeed, if we were only interested in the evolution of triple systems under Scenario C evolution, we could probably conclude the case here with the result that no Ba stars result from Scenario C evolution. However, it should be noted that, for many accretion scenarios where the mass loss ratio is in between 0 and 1 instead of exactly 0, angular momentum can be carried away by the lost mass, and the remnant system is not one in which angular momentum is conserved. In these cases, our assumption that all transferred mass is accreted by the inner binary while retaining its original angular momentum can be unphysical itself. To illustrate this point, consider the case in which the material forms a disk around the inner binary, which would be typical for a system evolving in between the ``disk + complete ejection" and ``disk + complete accretion" scenarios. If the material's angular momentum is retained, the disk cannot collapse onto the accreting stars and be accreted; it needs to first lose part of its angular momentum, possibly by ejecting a portion of its mass, in order to achieve this. To understand these systems in the context of them lying in between Scenario C and the other extreme scenarios, we also need to ask what happens when angular momentum is not conserved during Scenario C evolution.

Hence, we next consider the assumption that the specific angular momentum carried by the accreted material is the same as that of the material at the L1 Lagrange point of the outer orbit. This corresponds to the treatment outlined in Eq. \ref{leigh3}. The result of this scenario is that the amount of angular momentum injected by the accreted material is systematically many orders of magnitude greater than that of the inner binary, and the inner binary will hence unbind. No inner binaries of any of the 154 systems are driven closer, and it is even less likely to result in Ba stars than the earlier assumption that angular momentum is conserved throughout the triple system.

Finally, we pose the question of how many Ba stars can form if the infalling material carries no angular momentum whatsoever. This is unlikely to be the case for any given accretion process, but since all our other assumptions lead to too great an amount of angular momentum being introduced into the inner binary for a Ba star to form, it is useful to place an upper limit of sorts on how many Ba stars can form by assuming that the accreted material holds no angular momentum at all. This treatment corresponds to Eq. \ref{zero1}, and its results are plotted in Fig. \ref{scenarioC}. Without TTs, this results in 33 Ba stars, while 39 Ba stars are formed if TTs are considered, corresponding to a Galactic rate of $9.8{\times}10^{-5}$/yr and $1.2{\times}10^{-4}$/yr respectively. These results are summarised in Tab. \ref{results}, with the numbers quoted covering the entire range of all the results arrived at from the various different assumptions explored above.

\subsection{Scenario D Evolution}

For the situation in which all the mass from the tertiary is all accreted and then re-ejected, which we have dubbed Scenario D evolution, we calculate only the final distribution of $a_{\rm RL}/a_{\rm 1}$ under the assumption that infalling material holds no angular momentum, as the two other possibilities can lead to an unbinding of the inner binary prior to the re-ejection process. This final distribution of $a_{\rm RL}/a_{\rm 1}$ is provided in Fig. \ref{scenarioD}, where 14 and 20 Ba stars are formed without and with TTs taken into consideration, respectively. In terms of Galactic rates, this corresponds to $4.1{\times}10^{-5}$/yr and $5.9{\times}10^{-5}$/yr. These results are also summarised in Tab. \ref{results}, where we again note that these results are upper limits obtained under the assumption that accreted material holds no angular momentum.

\begin{table*}
	\centering
	\caption{Number of inner binaries that end up in Roche Lobe overflow in our different scenarios, both considering and not considering the effects of tertiary tides. The numbers  in the parentheses correspond to inferred Galactic rates.}
	\label{results}
	\begin{tabular}{ccccc}
		\hline
		Scenario & A & B & C & D  \\
		\hline
                no TTs & 10 ($3.0{\times}10^{-5}$/yr) & 99 ($2.9{\times}10^{-4}$/yr) & 0-33 ($9.8{\times}10^{-5}$/yr) & 0-14 ($4.1{\times}10^{-5}$/yr)  \\
                with TTs & 17 ($5.0{\times}10^{-5}$/yr) & 106 ($3.1{\times}10^{-4}$/yr) & 0-39 ($1.2{\times}10^{-4}$/yr) & 0-20 ($5.9{\times}10^{-5}$/yr)  \\
		\hline
	\end{tabular}
\end{table*}

\subsection{Outer Orbital Period Distribution}

To investigate whether Ba stars originating from our evolution channel can account for the 1000 day period gap, we attempt to place constraints on the distribution of post-RLOF outer orbital periods for all 154 hierarchical triple systems in our sample. To do this, we find the upper and lower limits of the final outer orbital periods for each of the 154 systems by means of the methods below.

For each system, the maximum and minimum values for $P_{\rm final}$ must either be found when ${\beta}=0$, in which case Eq. \ref{outP4} applies, ${\beta}=1$, in which case Eq. \ref{outP3} applies, or when ${\beta}$ is some value in between $0$ and $1$, in which case the first derivative of the right hand side of the lower equation in Eq. \ref{outP5} to ${\beta}$ must be equal to zero:
\begin{equation}
\frac{\rm d}{{\rm d}{\beta}}\left(\frac{P_{\rm final}}{P_{\rm init}}\right)=0,
\label{outPderiv1}
\end{equation}
\noindent where $P_{\rm final}/P_{\rm init}$ as a function of ${\beta}$ is given by Eq. \ref{outP5}. 

Numerically solving this equation, we find no solutions for any of our 154 systems other than ${\beta}=0$. Therefore, calculating the orbital periods of the outer orbits for all 154 hierarchical triple systems in our sample according to Eqs. \ref{outP3} and \ref{outP4} should yield the upper and lower limits for the final orbital periods of the resultant Ba star binaries.

\begin{figure}
\includegraphics[scale=0.25, angle=0, trim= 2cm 0cm 0cm 0cm]{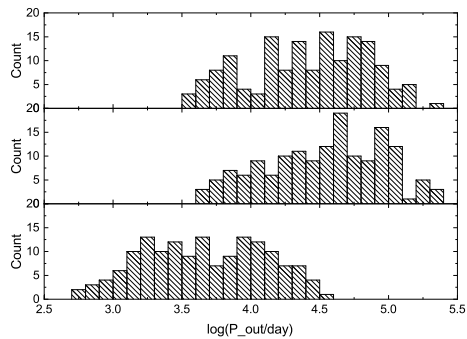}
\caption{Initial distribution and final expected distributions for the orbital periods of the outer orbits for our sample of 154 systems. Upper panel: distribution of the lower limits of the final outer orbital periods. Middle panel: distribution of the upper limits of the final outer orbital periods. Lower panel: Initial distribution of outer orbital periods for our sample of 154 hierarchical triple systems, which are candidates for producing Ba stars via the channel we investigate. \label{a2plots}}
\end{figure}

The final orbital periods of the systems containing our Ba stars, corresponding to the final outer orbits of our original triple systems, are plotted in Fig. \ref{a2plots}. The upper and lower limits for the orbital periods are plotted in the middle and upper panels respectively. For comparison, we also plot the initial outer orbital periods for our sample in the lower panel.

\section{Discussion}

In our studies, we find that hierarchical triple systems, which are able to undergo tertiary AGB mass transfer as stipulated in our model, are expected to be capable of producing a significant amount of Ba stars under a vast majority of the evolution channels that we have studied. The only evolution channels that might contribute little to overall Ba star rates are those in which much of the mass transferred from the tertiary is accreted prior to losing angular momentum, thereby preventing an inner binary merger. However, unless we have reason to believe that other channels, such as that of an inner binary CE, do not occur, it is inevitable that a large number of Ba stars are the result of hierarchical triple evolution. The highest formation rate inferred from our 4 evolutionary scenarios is found in Scenario B after TTs are considered, or in other words energy-inefficient full ejection of transferred mass by the inner binary, with tertiary tides accounted for under the assumption of resonant locking. This corresponds to a Galactic Ba star formation rate of $3.1{\times}10^{-4}$/yr. This, however, should not be taken to be an absolute upper limit, as we have disregarded those systems which form a common envelope and fail to eject it.

How does this compare with the actual Galactic Ba star formation rate? Given that previous studies have found that this formation rate is very sensitive to model assumptions \citep{1995MNRAS.277.1443H}, most notably the degree to which tidal forces enhance mass transfer rates, the answer could lie anywhere in between $0$ and $6.6{\times}10^{-3}$/yr. However, comparing their predictions of the number of observable Ba stars for each scenario with the 546 actually seen at a magnitude limit of 12 \citep{2017A&A...608A.100E}, it would appear that the real-world Galactic Ba star formation rate is in the neighbourhood of $6{\times}10^{-4}$/yr. If it is true that the scenarios in our study account for $3.1{\times}10^{-4}$ Ba stars per year (the Scenario B figures given above) then this would already account for roughly half the Galactic rate - and this is only for the triple formation channels that we have explored in this paper. 


In terms of their impact on Ba star formation rates, tertiary tides are found to play a certain, albeit limited, role in the creation of Ba stars in hierarchical triple systems. Judging from the results in Tab. \ref{results}, TTs may account for an enrichment of anything from 7\% to 70\% of the total number of Ba stars produced from the channels we have investigated, depending on the specific formation scenario. However, these numbers were obtained under the assumption that resonant locking systematically induces a $\tau$ value of $10^{-4}$ years during the RGB contraction phase. Whether this is reasonable for triple systems in general, however, remains to be seen. To this end, further studies on how tertiary tides behave for different systems are required.

Judging by the fact that no systems in our sample end up with values of $a_{\rm 2}$ that allow outer orbital periods in the 1000 day range, it appears that our model does not account for the 1000 day period gap predicted to exist among Ba stars. This is largely due to most of the outer orbital periods in our triple sample lying close to the upper bound of the period gap to begin with, coupled with stable mass transfer leading to the outer orbital periods increasing. This implies that other explanations may need to be invoked in order to account for this period gap, which the Ba star binaries that our sample produce do not cover. 

However, we note that many of our simplifying assumptions, may have played a role in introducing errors to the final outer orbital period distributions. For example, overlooking the effects of wind mass loss, which we do not consider, may have led to wider outer orbits than what physically is the case, as the mass lost from the system is expected to drain angular momentum; conversely, circularising all orbits prior to performing our calculations, which was done because many of our analytical methods are not compatible with eccentric orbits, must essentially increase the critical semi-major axis at which tertiaries can exist, thus leading to tighter outer orbits in our final distribution. To estimate the magnitude of these errors, we compared the results obtained using our calculations to those derived from the stellar evolution code SeBa (which does not make our simplifying assumptions) for 5 systems, and find that our final values for the orbital periods are sytematically larger by a factor of a few. Whether or not this uncertainty can end up with some of our systems lying inside the 1000 day period gap is beyond the scope of this study.

Here, it should also be noted that triple systems in which the outer orbit is much more compact do exist, and have been observed \citep[e.g.][]{2011Sci...331..562C}, and the main reason we do not investigate such systems is largely due to the uncertainty pertaining to the circumtriple common envelopes that must arise from such systems. Furthermore, due to our lack of knowledge of the physical processes that take place in triple systems, it is possible that the present study neglects hitherto unknown effects that might shrink the outer orbit. This is especially true when one considers the fact that the equations by which we calculate the final outer orbital period is essentially one that is used for binary orbits under non-conservative mass transfer, and thereby completely ignores the spin and torque created by the inner orbit, which may disproportionately influence the outer orbit under certain resonances. Whether these neglected aspects will have an impact of the final orbital separations for the Ba stars remains to be seen.

Another possibility by which hierarchical triple systems can produce Ba star binaries that lie within the 1000 day period gap is the case in which the inner binary fails to merge. In this case, all the mass donated by the tertiary ends up on one of the inner binary components, and the inner binary is consequently identified as a Ba star binary, the tertiary being too faint to be seen after being stripped of its envelope.

In terms of observational tests of our models, one may naively expect that Ba stars that have formed as a result of a merger of two stars would have a much faster spin than conventional Ba stars from binaries, as a result of conservation of angular momentum. However, recent studies seem to indicate that this is not always the case \citep[][e.g]{2016MNRAS.457.2355S}, as magnetic braking can slow down the spin in at least some situations. As such, other tests for our evolutionary channel will be necessary.

One issue that we have not been able to address in this paper is that of chemical mixing and its influence on observed surface abundances. This should be looked at more carefully in a future study. Since, for example, if the timescales for processes such as elemental diffusion and radiative levitation are sufficiently short relative to the age of the star, there is potential to constrain the properties of the donor at the time of mass transfer.

Should our hypothesis hold true that RLOF from an AGB tertiary can result in the inner binary becoming a Ba star once they merge, it is natural to deduce that we should be able to see the results of the remnant system if it fails to merge. At first glance, such a remnant system should consist of a close binary system of two Ba stars - not a very common sight - as its tertiary would likely take the form of a white dwarf, which may not easily be seen. However, upon closer inspection, one immediately realises that not all of these remnants would take such a form - many inner binaries which accrete the Ba-rich material via an accretion disk, for instance, are likely to have a dominating proportion of the accreted material landing on only the less massive star, resulting in a Ba star in orbit with a pre-AGB close companion. In fact, the discovery of such a Ba star with a pre-AGB companion is likely to be a smoking gun for the formation channel herein described, as there is no way for such a system to evolve from a binary. Similarly, the discovery of a W Uma binary with a nearby stripped tertiary, which would greatly support the narrative that inner binaries can receive mass from their tertiary companions and survive the experience, would also support the existence of binaries containing Ba stars that formed in the same way. As for other failed binary Ba star systems that may have arisen from hierarchical triple evolution, we note that double blue straggler binaries have already been observed \citep{2009Natur.462.1032M}. Regarding whether the inner binaries of a hierarchical triple system can merge as a result of interaction with the tertiary, there is already evidence hinting that some blue stragglers form as a result of merger events, at least in globular clusters \citep{2011MNRAS.410.2370L,2011MNRAS.416.1410L,2018NatAs...2..362S,2022MNRAS.509.3724R}.

To better understand Ba stars and the binaries in which they reside, it is important that we fully investigate how triple systems can give rise to these objects. Future work ought to pay more attention to observationally determining whether individual Ba stars are the products of inner binary merger events, as well as on finding physical processes that may be at work in triple systems, which may influence the results of the findings hitherto presented. To these ends of the latter, much work is already underway, which will be the contents of a future publication.




\section*{Acknowledgements}

YG is a Royal Society K.C. Wong International Fellow, and as such acknowledges funding from the Royal Society and the K.C. Wong Education Foundation.

ST acknowledges support from the Netherlands Research Council NWO (VENI 639.041.645 and VIDI 203.061 grants).

NWCL gratefully acknowledges the generous support of a Fondecyt Iniciaci\'on grant
11180005, as well as support from Millenium Nucleus NCN19-058 (TITANs) and funding
via the BASAL Centro de Excelencia en Astrofisica y Tecnologias Afines (CATA) grant
PFB-06/2007.  NWCL also thanks support from ANID BASAL project ACE210002 and ANID
BASAL projects ACE210002 and FB210003.

\section*{Data Availability Statement}

The data underlying this article will be shared on reasonable request to the corresponding author.








\appendix

\section{Details Regarding Tertiary Accretion}
\label{appA}

One of the prescriptions for angular momentum, for material that is being accreted, which we investigate, is
\begin{equation}
v_{\rm orb,3}a_{\rm 2}\left(1-R_{\rm L}\right)=v_{\rm circ}a_{\rm circ},
\end{equation}
\noindent where $v_{\rm orb,3}$ is the orbital velocity of the L1 Lagrange point of the tertiary relative to the inner binary's COM, $v_{\rm circ}$ and $a_{\rm circ}$ are, respectively, the circular orbital velocity and orbital separation of the accreted matter relative to the inner binary's COM.

Implicit in this prescription is the assumption that material being dumped from the tertiary onto the inner binary is at rest relative to the L1 Lagrange point while passing it. While not physically true, it is usually seen as small by many authors (e.g. \citealt{1975ApJ...198..383L}).

Strictly speaking, the $a_{\rm 2}\left(1-R_{\rm L}\right)$ factor on the left hand side of the equation should be replaced by the distance from the inner binary's COM to the L1 Lagrange point. However, for an order-of-magnitude estimate, this factor is precise enough, for reasons that we mention below.

Lastly, it should be noted that this equation does not ascertain angular momentum conservation within the hierarchical triple system. While unphysical, it should be noted that, in certain situations, mass loss can extract angular momentum from the system to facilitate accretion.

For one of our subsequent equations, Eq. \ref {leigh3}, repeated here:
\begin{equation}
\begin{split}
|{\bf J}_{\rm f}-{\bf J}|=&(m_{\rm 3}-m_{\rm c})\left(1-R_{\rm L}\right)^{\rm 2}\\
&\left[Ga_{\rm 2}(m_{\rm 1}+m_{\rm 2}+m_{\rm 3})\right]^{\frac{1}{2}},
\end{split}
\end{equation}
\noindent its true form should, in fact, be 
\begin{equation}
\begin{split}
|{\bf J}_{\rm f}-{\bf J}|=&{\int}^{m_{\rm 3}-m_{\rm c}}_{0} \left(1-R_{\rm L}\right)^{\rm 2}\\
&\left[Ga_{\rm 2}(m)(m_{\rm 1}+m_{\rm 2}+m_{\rm 3}-m)\right]^{\frac{1}{2}}dm,
\end{split}
\end{equation}
\noindent where $a_{\rm 2}(m)$ is the varying value of $a_{\rm 2}$ as it changes with increasing $m$. However, since an order-of-magnitude estimate is all that we need, we opt not to calculate this full integration. A similar simplification is used for Eq. \ref {cons1}.

The reason that we are satisfied with an order-of-magnitude estimate is that this prescription leads to the final orbital angular momentum being systematically 5 orders of magnitude higher than that of the original inner binary. As such, all our results hold whether or not a more precise version of our calculation is used. Incidentally, it should be pointed out that, if our results concerning this prescription were not the case, it would not detract from our conclusion that Ba stars can be formed from hierarchical triples.


\bsp	
\label{lastpage}
\end{document}